\documentclass[10pt,conference]{IEEEtran}
\IEEEoverridecommandlockouts
\usepackage{cite}
\usepackage{amsmath,amssymb,amsfonts}
\usepackage[noend]{algorithmic}
\usepackage{graphicx}
\usepackage{textcomp}
\usepackage{xcolor}
\usepackage{amssymb}
\usepackage{indentfirst}
\usepackage{amsmath}
\usepackage{autobreak}
\usepackage{bm}
\usepackage{booktabs}
\usepackage[font=small]{caption}
\usepackage{subcaption}

\usepackage{tikz}
\usepackage{pgfplots}
\usetikzlibrary{plotmarks}
\usetikzlibrary{matrix,positioning,shapes,arrows,decorations,calc,fit}
\usetikzlibrary{decorations.pathreplacing}
\usetikzlibrary{spy,backgrounds}
\usetikzlibrary{external}
\usetikzlibrary{patterns}
\usetikzlibrary{shapes,arrows}
\usetikzlibrary{shadows.blur}
\usetikzlibrary{shapes.symbols}
\usetikzlibrary{
	pgfplots.groupplots,
	matrix
}

\usepackage{epsfig}
\usepackage{setspace}
\usepackage{algorithm}
\usepackage{changepage}
\usepackage{amsthm} 
\usepackage{soul,color}
\usepackage{hyperref}
\usepackage{lipsum}
\usepackage{cuted}
\usepackage{multirow}
\newtheorem{proposition}{Proposition}

\theoremstyle{definition}

\IEEEaftertitletext{\vspace{-3\baselineskip}}

\begin{document}

\setlength{\abovedisplayskip}{3pt}
\setlength{\belowdisplayskip}{3pt}

\title{NOMA Joint Decoding based on Soft-Output Ordered-Statistics Decoder for Short Block Codes}
\author{\IEEEauthorblockN{
       Chentao Yue\IEEEauthorrefmark{1}, Alva Kosasih\IEEEauthorrefmark{1}, 
		Mahyar Shirvanimoghaddam\IEEEauthorrefmark{1}, Giyoon Park\IEEEauthorrefmark{2}, Ok-Sun Park\IEEEauthorrefmark{2},\\ Wibowo Hardjawana\IEEEauthorrefmark{1}, Branka Vucetic\IEEEauthorrefmark{1}, and Yonghui Li\IEEEauthorrefmark{1}}
\IEEEauthorblockA{\IEEEauthorrefmark{1}School of Electrical and Information Engineering, The University of Sydney, NSW, Australia \\
\{chentao.yue, alva.kosasih, mahyar.shm, wibowo.hardjawana, branka.vucetic, yonghui.li\}@sydney.edu.au}

\IEEEauthorblockA{\IEEEauthorrefmark{2}Electronics and Telecommunications Research Institute,
Daejeon, South Korea\\
\{gypark, ospark\}@etri.re.kr}

}

\maketitle

\begin{abstract}
In this paper, we design the joint decoding (JD) of non-orthogonal multiple access (NOMA) systems employing short block length codes. We first proposed a low-complexity soft-output ordered-statistics decoding (LC-SOSD) based on a decoding stopping condition, derived from approximations of the \textit{a-posterior} probabilities of codeword estimates. Simulation results show that LC-SOSD has the similar mutual information transform property to the original SOSD with a significantly reduced complexity. Then, based on the analysis, an efficient JD receiver which combines the parallel interference cancellation (PIC) and the proposed LC-SOSD is developed for NOMA systems. Two novel techniques, namely decoding switch (DS) and decoding combiner (DC), are introduced to accelerate the convergence speed. Simulation results show that the proposed receiver can achieve a lower bit-error rate (BER) compared to the successive interference cancellation (SIC) decoding over the additive-white-Gaussian-noise (AWGN) and fading channel, with a lower complexity in terms of the number of decoding iterations.

\end{abstract}

\IEEEpeerreviewmaketitle

\vspace{-0.2em}
\section{Introduction}
\vspace{-0.2em}

   Ultra-reliable and low-latency communications (URLLC) have attracted great attention in 5G and upcoming 6G for mission-critical services \cite{Mahyar2019ShortCode,popovski2019wireless}. Ultra-low latency requires low complexity receivers and mandates the use of short block-length codes ($\leq 150$ bits) \cite{Mahyar2019ShortCode}. Also, the scalable and reliable connectivity for a large number of users with limited channel spectrum resources is required for mission-critical services \cite{popovski2019wireless}. Non-orthogonal multiple access (NOMA) has recently gained popularity as a promising technique for improving spectral efficiency \cite{makki2020survey}. It allows users to transmit signals that are non-orthogonal in terms of frequency, time, or code domains in a superposed manner. The superposed signals can be detected using the successive interference cancellation (SIC)\cite{wang2004wireless}. NOMA can achieve certain corner points of the multiple-access channel (MAC) capacity region using SIC in the asymptotically large block length scenario \cite{wang2019near}. However, SIC is insufficient for URLLC applications due to its sequential nature. Specifically, the last decoded user has the worst latency, whereas the first decoded user faces severe multiple-access interference (MAI). NOMA should use a low-complexity joint decoding (JD) instead of SIC when providing URLLC services.
    
    The complexity of the maximum-likelihood (ML) JD of multi-user transmission grows exponentially with the number of users and codebook size. Many low-complexity JD schemes have been proposed in \cite{liu2019capacity,ping2004approaching,wang2019near,ebada2020iterative,sharifi2015ldpc,xiang2021iterative,zhang2017channel,balatsoukas2018design} for asymptotically large block length scenarios. They typically combine a multi-user detector (MUD) and an \textit{a-posterior} probability (APP) decoder, to iteratively perform MAI cancellation and decoding. With this iterative structure, LDPC codes with large length were analyzed and optimized for NOMA with belief propagation (BP) decoding \cite{wang2019near,sharifi2015ldpc,balatsoukas2018design}. Receiver designs with moderate/long polar codes were investigated in \cite{ebada2020iterative,xiang2021iterative} using BP or successive cancellation list (SCL) decoding. Although notable studies have been made for moderate/long block codes, designing practical NOMA JD receivers for short block length codes is rarely attempted in literature.
    
    Two critical factors should be carefully considered in the short block length regime: 1) the coding scheme and 2) the decoding algorithm. Most of the conventional powerful codes, including LDPC and Polar codes, fall short under the short block length regime when compared to the normal approximation (NA) bound \cite{liva2016codeSurvey,PPV2010l}. The short BCH codes have gained interest from the research community recently \cite{liva2016codeSurvey,Chentao2019SDD, yue2021probability}. BCH codes outperform other existing short codes in terms of block-error-rate performance and approximately approach NA, but its ML decoding is highly complex. As a universal near-ML decoder, the ordered-statistics decoding (OSD) rekindled interests \cite{Fossorier1995OSD,dhakal2016error, yue2021revisit, yue2021probability} in decoding high-density codes like BCH codes. In \cite{fossorier1998soft}, OSD was modified to output the posterior log-likelihood ratio (LLR) of codeword bits, referred to as soft-output OSD (SOSD). SOSD with high decoding order approximates the Max-Log-MAP algorithm \cite{hagenauer1996iterative}.
    
    In this paper, we design an iterative JD receiver for power-domain NOMA systems with short block codes based on SOSD. We first propose a low-complexity SOSD (LC-SOSD). We show that APPs of codeword estimates in SOSD can be approximated by a so-called success probability (SP). Then, SOSD can be terminated early when SP satisfies a certain threshold. Simulations show that LC-SOSD has the similar mutual information (MI) transform as the original SOSD but with less complexity. Next, an iterative JD receiver is devised by combining parallel interference canceller (PIC) \cite{kosasih2021bayesian} and LC-SOSD. Two techniques, decoding switch (DS) and decoding combiner (DC), are introduced to improve the proposed receiver. DS regulates LC-SOSD participation at early receiving iterations when MAI is high. DC, on the other hand, adaptively combines the decoder input and output based on a predefined decoding quality. Comparisons are made between the proposed JD and SIC for decoding short BCH codes over additive-white-Gaussian-noise (AWGN) and fading channels. It is shown that the proposed scheme achieves a lower bit-error rate (BER) than SIC decoding, while having fewer decoding iterations and a lower decoding complexity per iteration.
    
    The rest of this paper is organized as follows. Section \ref{sec::Preliminaries} introduces the system model. Section \ref{Sec::LowSISO} and \ref{Sec::Receiver} discuss LC-SOSD and the proposed JD Receiver, respectively. Section \ref{Sec::Simulation} presents the simulation results. Section \ref{sec::Conclusion} concludes the paper.
    
    \noindent\emph{Notation}: We use $\mathrm{Pr}(\cdot)$ to denote the probability of an event. We use $[a]_u^v = [a_u,\ldots,a_v]$ to denote a row vector containing element $a_{\ell}$ for $u\le \ell\le v$. $\mathcal{O}(\cdot)$ is the big-O operand.

\vspace{-0.2em} 
\section{Preliminaries} \label{sec::Preliminaries}
\vspace{-0.2em} 

\subsection{System Model}
\vspace{-0.3em}

    We consider a binary phase shift keying (BPSK) signal transmission over a block-fading channel in uplink NOMA with $N_u$ simultaneous users. Given generator matrix $\mathbf{G}$ of code ${\mathcal C}(n,k)$, the information block of user $u$, $\mathbf{b}^{(u)}$, is encoded to the codeword, $\mathbf{c}^{(u)}$, with $\mathbf{c}^{(u)} = \mathbf{b}^{(u)}\mathbf{G}$, where $k$ and $n$ denote the information block and codeword lengths, respectively. Generator matrix $\mathbf{G}$ is identical for all users. The codeword $\mathbf{c}^{(u)}$ is interleaved by a random interleaver $\Pi_u$. All users simultaneously transmit the modulated symbol to the base station non-orthogonally. At the base station, the superposed signal $\mathbf{r}$ is received as
        \begin{equation} \label{equ::Pri::Sysmod}
            \mathbf{r} = \mathbf{h}\mathbf{X} + \mathbf{w},
        \end{equation}
    where $\mathbf{h} = [h^{(1)},\ldots,h^{(N_u)}]$ is the channel coefficient vector. Coefficient $h^{(u)}$ follows a scaled complex variable $h^{(u)}\sim \rho^{(u)}\mathcal{CN}(0,1)$, where $(\rho^{(u)})^2$ is the average receiving power of user $u$. $\mathbf{X} = [\mathbf{x}^{(1)};\mathbf{x}^{(2)};\ldots;\mathbf{x}^{(N_u)}]$ is a $N_u\times n$ matrix of modulated symbols, where $\mathbf{x}^{(u)}$ is the symbol vector of $\mathbf{c}^{(u)}$, i.e., $x_{i}^{(u)} = (-1)^{c_{i}^{(u)}}$ for $1\leq i\leq n$. $\mathbf{w} = [w]_1^n$ is the independent AWGN vector, where $w_i \sim \mathcal{CN}(0,\sigma^2)$. At the receiver, we assume that the channel coefficients are known a priori, and define the multi-user SNR as $\mathrm{SNR} = \sum_{u=1}^{N_u}\frac{1}{\sigma^2}(\rho^{(u)})^2$.
    
\begin{figure} 
		\begin{center}
			\includegraphics[scale=0.52] {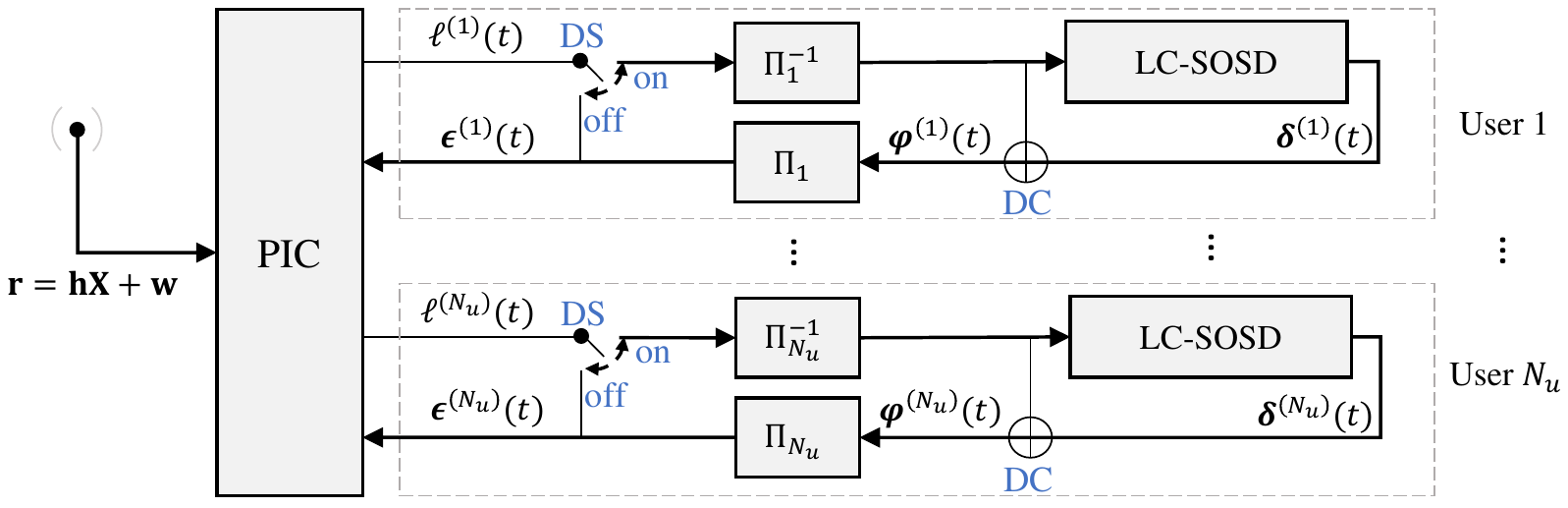}
	        \vspace{-0.31em}
			\caption{The structure of the proposed iterative JD receiver.}
	        \vspace{-0.81em}
			\label{Fig::Structure}
		\end{center}
	\end{figure}

    The signal $\mathbf{r}$ is received by the iterative JD receiver as shown in Fig \ref{Fig::Structure}. For clarity of notation, we do not differentiate variables before and after interleavers. We note that inteleavers are effective in  reducing the correlation between the signals of different users \cite{ping2003interleave,wang2019near}. The JD receiver has two major phases. First, DS is off at the beginning of JD. PIC finds the extrinsic LLRs, $\bm{\ell}^{(u)}(t) = [\ell^{(u)}(t)]_1^n$, where $t$ is the iteration index. Then, $\bm{\epsilon}^{(u)}(t) \leftarrow \bm{\ell}^{(u)}(t)$ is directly fedback to PIC, serving as the prior LLRs for the next iteration. Second, DS is turned on after a few iterations. $\bm{\ell}^{(u)}(t)$ is input to LC-SOSD to output $\bm{\delta}^{(u)}(t)$. Then, DC combines $\bm{\delta}^{(u)}(t)$ and $\bm{\ell}^{(u)}(t)$ to obtain $\bm{\varphi}^{(u)}(t)$ according to the decoding quality. Finally, $\bm{\epsilon}^{(u)}(t) \leftarrow \bm{\varphi}^{(u)}(t)$ is fedback to PIC for the next iteration. 
    
	\vspace{-0.3em}
	\subsection{Ordered-Statistics Decoding} \label{sec::Pri::OSD}
     \vspace{-0.3em} 
        We briefly introduce the OSD algorithm as follows. A sequence of LLR $\bm{\ell} = [\ell]_1^n$ of the transmitted codeword $\mathbf{c}$ is input to OSD, defined as ${\ell}_{i} \triangleq \log \frac{\mathrm{Pr}(c_{i}=1|\bar{\mathbf{x}})}{\mathrm{Pr}(c_{i}=0|\bar{\mathbf{x}})}$ conditioning on a observation $\bar{\mathbf{x}}$ of $\mathbf{c}$. Starting OSD, the bit-wise hard-decision estimate $\mathbf{y}= [y]_{1}^n$ is first obtained based on $\bm{\ell}$ according to $y_{i}= 1$ for $\ell_{i}<0$ and $y_{i}= 0$ for $\ell_{i}\geq 0$. We define the magnitude of LLR $\ell_{i}$ as the reliability of $y_{i}$, denoted by $\alpha_{i} = |\ell_{i}|$, where $|\cdot|$ is the absolute operation. 
    	
        Then, a permutation $\pi_1$ is performed to sort $\bm{\ell}$ and columns of $\mathbf{G}$ in the descending order of reliabilities $\bm{\alpha}$.  Next, Gaussian elimination (GE) is performed to obtain the systematic form of permuted matrix $\pi_1(\mathbf{G})$, i.e., $\mathbf{\widetilde{G}} = [\mathbf{I}_k \  \mathbf{\widetilde{P}}]$, where $\mathbf{I}_k$ is a $k\times k$ identity matrix and $\mathbf{\widetilde{P}}$ is the parity sub-matrix. An additional permutation $\pi_{2}$ may occur during GE to ensure that the first $k$ columns of $\mathbf{\widetilde G}$ are linearly independent. After all permutations, the input LLR, reliability, and the generator matrix are permuted to $\widetilde{\bm{\ell}} = \pi_{2}(\pi_{1}(\bm{\ell}))$, $\bm{\widetilde \alpha} = \pi_{2}(\pi_{1}(\bm\alpha))$, and $\mathbf{\widetilde G} = \pi_{2}(\pi_{1}(\mathbf{G}))$, respectively. As shown by \cite[Eq. (59)]{Fossorier1995OSD}, $\pi_{2}$ can be usually omitted. $\mathbf{\widetilde{y}}_{\mathrm{B}} = [\widetilde{y}]_1^k$ is referred to as the most reliable basis (MRB). Throughout the paper, we use subscript $\mathrm{B}$ and $\mathrm{P}$ to denote the first $k$ positions and the rest positions of a length-$n$ vector, e.g., $\mathbf{\widetilde{y}} = [\mathbf{\widetilde{y}}_{\mathrm{B}} \ \ \mathbf{\widetilde{y}}_{\mathrm{P}}]$.
    
        In OSD, a number of TEPs are checked to find the best codeword estimate. A codeword estimate $\mathbf{\widetilde c}_{\mathbf e}$ is generated by re-encoding a TEP $\mathbf{e} = [e]_1^k$ as follows: $\mathbf{\widetilde c}_{\mathbf{e}} = \left(\widetilde{\mathbf{y}}_{\mathrm{B}}\oplus \mathbf{e}\right)\mathbf{\widetilde G} = [\widetilde{\mathbf{y}}_{\mathrm{B}}\oplus \mathbf{e} \  ~\left(\widetilde{\mathbf{y}}_{\mathrm{B}}\oplus \mathbf{e}\right)\mathbf{\widetilde{P}}] $,
    	where $\mathbf{\widetilde c}_{\mathbf{e}} = [{\widetilde c}_{\mathbf{e}}]_1^n$ is the codeword estiamte with respect to $\mathbf{e}$. TEPs are checked in an increasing order of their Hamming weights. The maximum Hamming weight of the TEP will be limited, which is referred to as the decoding order. As shown in \cite{Fossorier1995OSD,yue2021revisit}, the overall complexity of OSD is highly determined by the size of TEP list.
    	
    	For BPSK modulation, finding the best ordered codeword estimation $\mathbf{\widetilde c}_{\mathrm{op}}$ is equivalent to minimizing the weighted Hamming distance (WHD) between $\widetilde{\mathbf{c}}_{\mathbf{e}}$ and $\mathbf{\widetilde{y}}$, which is defined as $d(\mathbf{\widetilde c}_{\mathbf{e}},\mathbf{\widetilde y}) \triangleq \sum_{\substack{0<i \leq n \\ \widetilde{c}_{\mathbf{e},i}\neq \widetilde{y}_{i}}} \widetilde\alpha_{i}$.
        We take $d_{\mathbf{e}} = d(\mathbf{\widetilde c}_{\mathbf{e}},\mathbf{\widetilde y})$ for simplicity. Finally, the optimal estimation $\hat{\mathbf{c}}_{\mathrm{op}}$ corresponding to the input LLR sequence $\bm{\ell}$, is obtained by performing inverse permutations over $\mathbf{\widetilde c}_{\mathrm{op}}$, i.e.
    	$\hat{\mathbf{c}}_{\mathrm{op}} = \pi_{1}^{-1}(\pi_{2}^{-1}(\mathbf{\widetilde c}_{\mathrm{op}}))$.
    
    	In \cite{fossorier1998soft}, OSD was modified to output soft information, which is referred to as the SOSD. Given the input LLR sequence $\bm{\ell}$, after decoding, the extrinsic LLR of the $\ell_i$ is derived as \cite{fossorier1998soft}
    	\begin{equation} \label{equ::LowSISO::outLLR2}
            \delta_{i} \! =\! \log\!\left(\!\frac{ \mathrm{Pr}(\mathbf{c}(i\!:\!0) \!=\! \mathbf{c}|\bm{\ell})}{ \mathrm{Pr}(\mathbf{c}(i\!:\!1)\! =\! \mathbf{c}|\bm{\ell})}\! \right) \!-\! \ell_i  \!=\!\! \sum_{\substack{1\leq j \leq n \\ j\neq i}}\!\!\ell_j\left(c_j(i\!:\!1) \!-\! c_j(i\!:\!0)\right) \!,
        \end{equation}
        where $\mathbf{c}(i:0)$ and $\mathbf{c}(i:1)$ are respectively the codeword estimate whose $i$-th bit is 0 and 1, with lowest WHD to $\mathbf{y}$.
	
	\vspace{-0.2em}
	\section{Low-complexity SOSD} \label{Sec::LowSISO}
	\vspace{-0.2em}
        In order to employ early decoding stopping conditions to reduce the complexity of SOSD, we can rewrite (\ref{equ::LowSISO::outLLR2}) as
        \begin{equation} \label{equ::LowSISO::transMAP}
            \delta_{i} = (-1)^{\hat{c}_{\mathrm{\mathrm{op}},i}}\log\left( \frac{ \mathrm{Pr}(\hat{\mathbf{c}}_{\mathrm{op}} = \mathbf{c} |\bm{\ell}) }{ \mathrm{Pr}(\mathbf{c}_{\mathrm{sub}}(i) = \mathbf{c} |\bm{\ell}) } \right) - \ell_i .\\
        \end{equation}
        We can see that $\mathrm{Pr}(\hat{\mathbf{c}}_{\mathrm{op}} = \mathbf{c} |\bm{\ell})$, is identical to any position $i$, $1 \leq i \leq n$, in the same transmitted block. On the other hand, $\mathbf{c}_{\mathrm{sub}}(i)$ denotes the most likely codeword in the set $\mathcal{C}_i^{(1 - \hat{c}_{\mathrm{\mathrm{op}},i})} = \{\mathbf{c}\in\mathcal{C}:c_i = 1 - \hat{c}_{\mathrm{\mathrm{op}},i}\}$, which is in fact one of the sub-optimal codewords in the codebook $\mathcal{C}(n,k)$. We note that $\mathbf{c}_{\mathrm{sub}}(i)$ might differ for different position $i$. The approach of LC-SOSD is to directly approximate the APPs $\mathrm{Pr}(\hat{\mathbf{c}}_{\mathrm{op}} = \mathbf{c} |\bm{\ell})$ and $\mathrm{Pr}(\mathbf{c}_{\mathrm{sub}}(i) = \mathbf{c} |\bm{\ell})$. Then, the decoding is terminated early when APPs satisfy some conditions.
        
        \vspace{-0.31em}
       \subsection{Approximation of the APPs of the Codeword Estimates}  \label{Sec::LowSISO::appAPP}  
       \vspace{-0.31em}
        In \cite{yue2021revisit}, a probability regarding a TEP $\mathbf{e}$, referred to as the success probability (SP), is used to evaluate the likelihood of estimate\footnote{There exist relationships $\widetilde{\mathbf{c}}_{\mathbf{e}} = \pi_2(\pi_1(\mathbf{c}_{\mathbf{e}}))$ and $\mathbf{c}_{\mathbf{e}} = \pi_2^{-1}(\pi_1^{-1}(\widetilde{\mathbf{c}}_{\mathbf{e}}))$. For simplicity, we refer to both $\widetilde{\mathbf{c}}_{\mathbf{e}}$ and  $\mathbf{c}_{\mathbf{e}}$ as codeword estimates.} $\widetilde{\mathbf{c}}_{\mathbf{e}}$. Let $\widetilde{\mathbf{e}} = [\widetilde{\mathbf{e}}_{\mathrm{B}} \ \  \widetilde{\mathbf{e}}_{\mathrm{P}}]$ denote the hard-decision error pattern over the vector $\widetilde{\mathbf{y}}$, i.e., $\widetilde{\mathbf{e}} = \widetilde{\mathbf{y}}\oplus  \widetilde{\mathbf{c}}$, where $\widetilde{\mathbf{c}}$ is the permuted transmitted codeword, i.e., $\widetilde{\mathbf{c}} = \pi_2(\pi_1(\mathbf{c}))$. Then, the SP is defined as a conditional probability $\mathrm{Pr}(\mathbf{e} = \widetilde{\mathbf{e}}_{\mathrm{B}}|\bm{\ell}, D_{\mathbf{e}} \!=\! d_{\mathbf{e}})$, where $D_{\mathbf{e}}$ is a random variable representing the WHD $d_{\mathbf{e}}$. Next, we prove that the SP of $\mathbf{e}$ is equivalent to the APP of the corresponding codeword estimate, i.e., $\mathrm{Pr}(\mathbf{c}_{\mathbf{e}} = \mathbf{c} |\bm{\ell})$. To begin, we have the following proposition.
        \begin{proposition} \label{Pro::LowSISI::Pcorrect}
             If the errors of MRB $\widetilde{\mathbf{e}}_{\mathrm{B}}$ is eliminated by a TEP $\mathbf{e}$, the corresponding codeword estimate $\mathbf{c}_{\mathbf{e}}$ of $\mathbf{e}$ is the transmitted codeword $\mathbf{c}$.
        \end{proposition}
            \begin{IEEEproof}
                There are maximum $2^k$ possible TEPs corresponding to $2^{k}$ different codewords of $\mathcal{C}(n,k)$. Also, the hard-decision vector $\widetilde{\mathbf{y}}$ can be represented as $\widetilde{\mathbf{y}} = \widetilde{\mathbf{c}} \oplus \widetilde{\mathbf{e}}$.
                
                If the errors of MRB $\widetilde{\mathbf{e}}_{\mathrm{B}}$ is \textbf{not} eliminated by a TEP $\mathbf{e}'$, the codeword estimate $\widetilde{\mathbf{c}}_{\mathbf{e}'}$ corresponding to $\mathbf{e}'$ can be re-written as $ \widetilde{\mathbf{c}}_{\mathbf{e}'} = [\widetilde{\mathbf{y}}_{\mathrm{B}}\oplus \mathbf{e}']\widetilde{\mathbf{G}} = [\widetilde{\mathbf{c}}_{\mathrm{B}} \oplus \widetilde{\mathbf{e}}_{\mathrm{B}} \oplus \mathbf{e}']\widetilde{\mathbf{G}}$.
                Since $\widetilde{\mathbf{e}}_{\mathrm{B}} \oplus \mathbf{e}' \neq \mathbf{0}$, $\widetilde{\mathbf{c}}_{\mathbf{e}'}$ is not the correct codeword estimate, i.e., $ \widetilde{\mathbf{c}}_{\mathbf{e}'} \neq \widetilde{\mathbf{c}}$. Furthermore, by considering that there are $2^k$ TEPs in total, there is only one TEP $\mathbf{e}$ that can eliminate the MRB errors $\widetilde{\mathbf{e}}_{\mathrm{B}}$. Thus, we can conclude that the codeword estimate $\widetilde{\mathbf{c}}_{\mathbf{e}}$ regarding $\mathbf{e}$ has to be the transmitted codeword.
            \end{IEEEproof}
        
        From Proposition \ref{Pro::LowSISI::Pcorrect}, we can directly conclude that $\mathrm{Pr}(\mathbf{e} = \widetilde{\mathbf{e}}_{\mathrm{B}}) = \mathrm{Pr}(\widetilde{\mathbf{c}}_{\mathbf{e}} = \widetilde{\mathbf{c}}) $. Then, we have the following proposition.
        
        \begin{proposition} \label{Pro::LowSISI::Equiv}
             For a codeword estimate $\widetilde{\mathbf{c}}_{\mathbf{e}}$ and its corresponding TEP $\mathbf{e}$, $\mathrm{Pr}(\mathbf{e} = \widetilde{\mathbf{e}}_{\mathrm{B}}|\bm{\ell}, D_{\mathbf{e}} \!=\! d_{\mathbf{e}}) = \mathrm{Pr}(\mathbf{c}_{\mathbf{e}} = \mathbf{c}|\bm{\ell})$.
        \end{proposition}
         Proposition \ref{Pro::LowSISI::Equiv} is immediately proved by noting that the condition $\{ D_{\mathbf{e}} \!=\! d_{\mathbf{e}}\}$ is held when $\bm{\ell}$ and $\mathbf{c}_{\mathbf{e}}$ are given according to the definition of WHD $d_{\mathbf{e}}$.
        
         Proposition \ref{Pro::LowSISI::Equiv} indicates that SP is equivalent to the APP of a codeword estimate. Furthermore, since $\widetilde{\bm{\ell}}$ determines $\widetilde{\mathbf{y}}$, we can rewrite $\mathrm{Pr}(\mathbf{e} = \widetilde{\mathbf{e}}_{\mathrm{B}}|\bm{\ell}, D_{\mathbf{e}} \!=\! d_{\mathbf{e}})$ as $\mathrm{Pr}(\mathbf{e}|\bm{\ell}, \widetilde{\mathbf{d}}_{\mathbf{e}})$, where $\widetilde{\mathbf{d}}_{\mathbf{e}} = \widetilde{\mathbf{c}}_{\mathbf{e}} \oplus \widetilde{\mathbf{y}}$ is the difference pattern between $\widetilde{\mathbf{c}}_{\mathbf{e}}$ and $\widetilde{\mathbf{y}}$. Using the approach in \cite[Corollary 6]{yue2021revisit}, the SP $\mathrm{Pr}(\mathbf{e}|\bm{\ell}, \widetilde{\mathbf{d}}_{\mathbf{e}})$ can be approximately derived as
         \begin{equation}  \label{equ::LowSISO::SPapp}
            \mathrm{Pr}(\mathbf{e}|\bm{\ell}, \widetilde{\mathbf{d}}_{\mathbf{e}}) \approx
          			\Bigg(1 \! + \! \frac{(1-\mathrm{P}(\mathbf{e}))2^{k-n}}{\mathrm{P}(\mathbf{e})\prod_{\substack{k < i \leq n\\\widetilde{d}_{\mathbf{e},i} \neq 0}} \mathrm{P}(i)\prod_{\substack{k < i \leq n\\ \widetilde{d}_{\mathbf{e},i}= 0 }} (1- \mathrm{P}(i))} \Bigg)^{\!-1} \!\!.
        \end{equation}
        where $\mathrm{P}(i)$ is the probability that the $i$-th bits of $\widetilde{\mathbf{y}}$ is in error conditioning on $\ell_i$, i.e.
        \begin{equation}             \label{equ:pei}
            \mathrm{P}(i) = (1+\exp(|\ell_i|))^{-1},
        \end{equation}
        and $\mathrm{P}(\mathbf{e}) $ is given by $
            \mathrm{P}(\mathbf{e})= \prod_{\substack{1 \leq i \leq k\\ e_i \neq 0}} \mathrm{P}(i)\prod_{\substack{1 \leq i \leq k\\ e_i = 0 }} (1- \mathrm{P}(i))$.
        The approximation in (\ref{equ::LowSISO::SPapp}) comes from assuming $\mathcal{C}(n,k)$ as a random code. By reusing $\mathrm{P}(i)$, (\ref{equ::LowSISO::SPapp}) is computed with $\mathcal{O}(n)$ multiplications. We omit the detailed derivation of (\ref{equ::LowSISO::SPapp}) due to space limits and refer readers to our previous work \cite{yue2021revisit}. 
        
         If SP can be computed precisely, $\mathrm{Pr}(\mathbf{e}|\bm{\ell}, \widetilde{\mathbf{d}}_{\mathbf{e}}) \geq 0.5$ indicates that  $\widetilde{\mathbf{c}}_{\mathbf{e}}$ must be the optimal codeword estimate, because APPs of all possible codeword estimates sums up to be 1. However, because (\ref{equ::LowSISO::SPapp}) only approximates SP, we need to introduce a parameter $\lambda$, $0.5\leq \lambda \leq 1$. Then, if $\mathrm{Pr}(\mathbf{e}|\bm{\ell}, \widetilde{\mathbf{d}}_{\mathbf{e}}) \geq \lambda$, $\widetilde{\mathbf{c}}_{\mathbf{e}}$ is claimed as the $\widetilde{\mathbf{c}}_{\mathrm{op}}$, and $\mathrm{Pr}(\hat{\mathbf{c}}_{\mathrm{op}} = \mathbf{c} |\bm{\ell})$ in (\ref{equ::LowSISO::transMAP}) is found accordingly. Unlike $\mathrm{Pr}(\hat{\mathbf{c}}_{\mathrm{op}} = \mathbf{c} |\bm{\ell})$, the APP of the sub-optimal codeword, i.e., $\mathrm{Pr}(\mathbf{c}_{\mathrm{sub}}(i) = \mathbf{c} |\bm{\ell})$ in (\ref{equ::LowSISO::transMAP}), is hardly early identified, because the sum of APPs of all codewords in $\mathcal{C}_i^{(1 - \hat{c}_{\mathrm{\mathrm{op}},i})}$ is unknown. Thus, we simply approximate $ \mathrm{Pr}(\mathbf{c}_{\mathrm{sub}}(i) = \mathbf{c} |\bm{\ell}) $ by the maximum SP among all generated codewords, whose $i$-th bit is opposite to $\hat{c}_{\mathrm{op},i}$. Later, we will show that this approximation only deviates the output marginally in terms of MI transform.
         
       \vspace{-0.31em} 
   \subsection{Algorithm of LC-SOSD}   
 \vspace{-0.31em}
 
        An order-$m$ LC-SOSD re-encodes the TEPs sequentially from $\mathbf{e}_{1}$ to $\mathbf{e}_{N_{\mathrm{max}}}$, where $N_{\mathrm{max}} = \sum_{j=0}^{m}\binom{k}{j}$. It calculates and stores the SP according to (\ref{equ::LowSISO::SPapp}) every time after re-encoding a TEP $\mathbf{e}$. Let $N_{a}$ denote the number of TEPs that have been re-encoded during the decoding, $1 \leq N_{a} \leq N_{\max}$. Then, computed SPs are stored in a set $\mathcal{P}$, i.e., $\mathcal{P} = \{\mathrm{Pr}(\mathbf{e}_1|\bm{\ell}, \widetilde{\mathbf{d}}_{\mathbf{e}_1}),\ldots,\mathrm{Pr}(\mathbf{e}_{N_{a}}|\bm{\ell}, \widetilde{\mathbf{d}}_{\mathbf{e}_{N_a}}) \}$. On the other hand, two length-$n$ lists, $\mathcal{P}^{(1)} = [\mathrm{P}_{1}^{(1)},\mathrm{P}_{2}^{(1)},\dots,\mathrm{P}_{n}^{(1)}]$ and $\mathcal{P}^{(0)}=[\mathrm{P}_{1}^{(0)},\mathrm{P}_{2}^{(0)},\dots,\mathrm{P}_{n}^{(0)}]$ are initialized with $\mathrm{P}_{i}^{(1)} = \mathrm{P}_{i}^{(0)} = 0$ for $1\leq i \leq n$. When a codeword estimate $\widetilde{\mathbf{c}}_{\mathbf{e}}$ is generated, $\mathrm{P}_{i}^{(\widetilde{c}_{\mathbf{e},i})}$ is updated to $\mathrm{P}_{i}^{(\widetilde{c}_{\mathbf{e},i})} = \mathrm{Pr}(\mathbf{e}|\bm{\ell}, \widetilde{\mathbf{d}}_{\mathbf{e}})$ if $\mathrm{Pr}(\mathbf{e}|\bm{\ell}, \widetilde{\mathbf{d}}_{\mathbf{e}})\geq \mathrm{P}_{i}^{(\widetilde{c}_{\mathbf{e},i})}$.
        
        Moreover, let $\mathrm{P}_{\max}$ denote the maximum entry in $\mathcal{P}$, i.e.,
        \begin{equation} \label{equ::Define::Pmax}
            \mathrm{P}_{\max} \triangleq \max\{\mathrm{Pr}(\mathbf{e}_i|\bm{\ell},\widetilde{\mathbf{d}}_{\mathbf{e}_i}) : 1\leq i \leq N_a\}.
        \end{equation}
        Then, for a predetermined $\tau$, $0.5\leq \tau \leq 1$, the LC-SOSD is terminated immediately if the following conditions are both satisfied: 1) $\mathrm{P}_{\max} \geq \tau$ and 2) $\min\{\mathcal{P}^{(1)}\} > 0$ and $\min\{\mathcal{P}^{(0)}\} > 0$. The second condition is for avoiding infinite values of posterior LLR. Upon the termination of decoding, the ordered extrinsic LLR $\widetilde{\delta}_{i}$ corresponding to $\widetilde{\ell}_i$ is obtained as
        \begin{equation} \label{Pro::LowSISI::EvaLLR}
            \widetilde{\delta}_{i} \approx (-1)^{\widetilde{c}_{\mathrm{\mathrm{op}},i}}\log\left( \frac{ \mathrm{P}_{\max} }{\mathrm{P}_{i}^{(1-\widetilde{c}_{\mathrm{op},i})}} \right) - \ell_i.
        \end{equation}
        Finally, the posterior LLRs are output by performing the inverse permutation of OSD, i.e., $\bm{\delta} = \pi_1^{-1}(\pi_2^{-1}(\widetilde{\bm{\delta}}))$.
        
        To further reduce the decoding complexity, we also integrate the TEPs discarding rule of \cite{yue2021probability}. Specifically, each TEP $\mathbf{e}$ is associated with a probability $\mathrm{P}_{p}(\mathbf{e})$ \cite[Eq. (8)]{yue2021probability} computed before re-encoding. Then, if $\mathrm{P}_{p}(\mathbf{e})$ is less than a threshold $\lambda_p$, the TEP $\mathbf{e}$ is skipped without re-encoding. This discarding rule can be implemented efficiently to reduce the complexity of OSD. We refer interested readers to \cite{yue2021probability}. The algorithm of the proposed LC-SOSD is summarized in Algorithm \ref{ago::LC-SISO}.

        \begin{algorithm} 

        	\caption{The proposed LC-SOSD}
        	\label{ago::LC-SISO}
        	\begin{algorithmic} [1]
        		\REQUIRE $\mathbf{G}$, $m$, $\bm{\ell}$, $\lambda_s$, and $\lambda_p$
        		\ENSURE The extrinsic LLR $\bm{\delta}$

        		\STATE Calculate reliability value $\alpha_{i} = |\ell_{i}|$
        		\STATE Hard decision: $y_{i}= 1$ for $\ell_{i}<0$ and $y_{i}= 0$, otherwise.
        		\STATE Permutations and GE to obtain $\widetilde{\bm{\alpha}}$, $\bm{\widetilde{\ell}}$,  $\bm{\widetilde{y}}$, and $\mathbf{\widetilde G}$
        		\STATE Initialize $\mathcal{P}^{(1)}$, $\mathcal{P}^{(0)}$ and $\mathrm{P}_{\max} = 0$

        		\FOR{$l=0:m$}
        		\FOR{$i=1:\binom{k}{l}$}
        		\STATE Take an new TEP $\mathbf{e}_{i}$ with Hamming weight $l$.
        		\STATE Compute $\mathrm{P}_{p}(\mathbf{e})$ according to \cite[Eq. (8)]{yue2021probability}
        		\STATE \textbf{if} $\mathrm{P}_{p}(\mathbf{e})\leq \lambda_p$ \textbf{then} \textbf{Continue}
        		\STATE Re-encoding $\mathbf{\widetilde c}_{\mathbf{e}_i} = \left(\mathbf{\widetilde y}_{\textup B}\oplus \mathbf{e}\right)\mathbf{\widetilde G}$ 
        		\STATE Compute $\mathrm{Pr}(\mathbf{e}_i|\bm{\ell}, \widetilde{\mathbf{d}}_{\mathbf{e}_i})$ according to (\ref{equ::LowSISO::SPapp}).
                \IF{$\mathrm{Pr}(\mathbf{e}_i|\bm{\ell}, \widetilde{\mathbf{d}}_{\mathbf{e}_i}) \geq \mathrm{P}_{\max}$}
        		    \STATE $\mathrm{P}_{\max} = \mathrm{Pr}(\mathbf{e}_i|\bm{\ell}, \widetilde{\mathbf{d}}_{\mathbf{e}_i}) $ and $\widetilde{\mathbf{c}}_{\mathrm{op}} = \mathbf{\widetilde c}_{\mathbf{e}_i} $
        		\ENDIF
        		\FOR{$j=1:n$}
        		    \STATE \textbf{if} $\mathrm{Pr}(\mathbf{e}|\bm{\ell}, \widetilde{\mathbf{d}}_{\mathbf{e}})\geq \mathrm{P}_{i}^{(\widetilde{c}_{\mathbf{e},i})}$  \textbf{then}  $\mathrm{P}_{i}^{(\widetilde{c}_{\mathbf{e},i})}= \mathrm{Pr}(\mathbf{e}|\bm{\ell}, \widetilde{\mathbf{d}}_{\mathbf{e}})$
        		    
        		\ENDFOR 
        		\IF{$\mathrm{P}_{\max} \geq \lambda_s$, $\min\{\mathcal{P}^{(1)}\}>0$, $\min\{\mathcal{P}^{(0)}\}>0$}
            		\STATE Go to step 19
        		\ENDIF
        		\ENDFOR		
        		\ENDFOR 
        		\STATE Calculate $\widetilde{\bm{\delta}} = [\widetilde{\delta}]_1^n$ according to (\ref{Pro::LowSISI::EvaLLR}).
        		\RETURN $\bm{\delta} = \pi_1^{-1}(\pi_2^{-1}(\widetilde{\bm{\delta}}))$
        	\end{algorithmic}
        \end{algorithm}

      \vspace{-0.3em}  
     \subsection{Comparison with the Original SOSD}  \label{Sec::LowSISO::Cmp}
        \vspace{-0.3em}
        
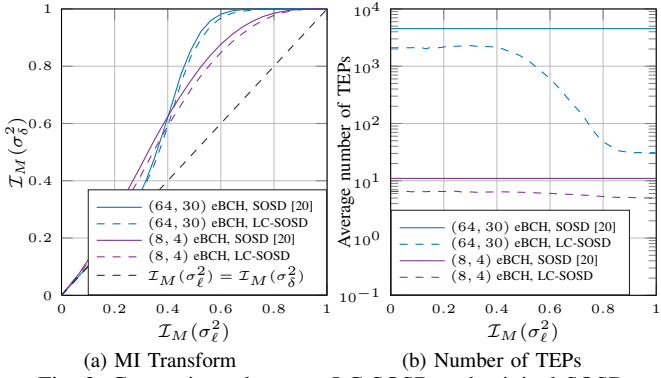
\begin{figure}
             \vspace{-0.25em}
             \centering
             \hspace{-0.81em}
             \begin{subfigure}[b]{0.49\columnwidth}
                 \centering
                    \definecolor{mycolor1}{rgb}{0.00000,0.44700,0.74100}%
                    \definecolor{mycolor2}{rgb}{0.00000,0.44706,0.74118}%
                    \definecolor{mycolor3}{rgb}{0.49412,0.18431,0.55686}%
                    \definecolor{mycolor4}{rgb}{0.14902,0.14902,0.14902}%
                    \begin{tikzpicture}
                    
                    \begin{axis}[%
                    width=1.39in,
                    height=1.5in,
                    at={(0,0.49in)},
                    scale only axis,
                    xmin=0.00,
                    xmax=1.00,
                    xlabel style={at={(0.5,2ex)},font=\color{white!15!black},font = \scriptsize},
                    xlabel={$\mathcal{I}_{M}(\sigma_{\ell}^2)$},
                    ymin=0,
                    ymax=1,
                    ylabel style={at={(4ex,0.5)},font=\color{white!15!black},font = \scriptsize},
                    ylabel={$\mathcal{I}_{M}(\sigma_{\delta}^2)$},
                    axis background/.style={fill=white},
                    tick label style={font=\tiny},
                    xmajorgrids,
                    ymajorgrids,
                    legend style={at={(1,0)}, anchor=south east, legend cell align=left, align=left, draw=white!15!black,font = \tiny,row sep=-3pt}
                    ]
                    \addplot [color=mycolor1]
                      table[row sep=crcr]{%
                    0	0\\
                    0.0696734432272046	0.0712052710510156\\
                    0.0767937327525715	0.0788944779179725\\
                    0.0862864787561703	0.0874993356404435\\
                    0.0952480659248293	0.0965271704675783\\
                    0.106109747649723	0.108448792444997\\
                    0.118196702844696	0.120744173755519\\
                    0.13117437236672	0.133570990933344\\
                    0.146135572854705	0.149009615690639\\
                    0.161725637176674	0.164863990271979\\
                    0.17891515630525	0.185141578610447\\
                    0.19865238952987	0.204824074547664\\
                    0.218210833235988	0.229560877693759\\
                    0.241146102520816	0.262456811743598\\
                    0.265112386609737	0.298415547380955\\
                    0.291615792010423	0.345224801812067\\
                    0.319824649024135	0.404620168066644\\
                    0.348498972574913	0.475132045393114\\
                    0.380298768816435	0.56346200883743\\
                    0.414408941525855	0.6635873653408\\
                    0.450433166615504	0.77308982577218\\
                    0.48628419314194	0.855014463367199\\
                    0.524867667738141	0.921242024542813\\
                    0.56277834024212	0.960523894504303\\
                    0.602066155784567	0.982458972294035\\
                    0.64229684127368	0.993389293360858\\
                    0.681505509912605	0.997658555524857\\
                    0.720952734563931	0.9993262865309\\
                    0.759066290609666	0.999835065443478\\
                    0.79442608336539	0.99996423433313\\
                    0.828027520572296	0.999993811577984\\
                    0.859099175109772	0.999999100035591\\
                    0.887230877719948	0.999999908288189\\
                    0.911654602638462	0.999999992286155\\
                    0.932898621186249	0.999999999564611\\
                    };
                    \addlegendentry{ $(64,30)$ eBCH, SOSD \cite{fossorier1998soft}}
                    
                    \addplot [color=mycolor2, dashed]
                      table[row sep=crcr]{%
                    0	0\\
                    0.0692738612653797	0.0704319281338104\\
                    0.0860376051482897	0.0877771449560154\\
                    0.106672584515217	0.108635656534146\\
                    0.130781351906411	0.133375417830153\\
                    0.161188156357534	0.165075261417479\\
                    0.197659419170211	0.204595574702169\\
                    0.241028217999545	0.256921187537786\\
                    0.290284480581853	0.333069283490755\\
                    0.347953984868309	0.453285246670595\\
                    0.380010268811711	0.534125253962762\\
                    0.414099522895813	0.631516318214965\\
                    0.450056273562446	0.742231697591567\\
                    0.486298366806477	0.83210975383399\\
                    0.524885213243445	0.895503973667289\\
                    0.562219816895679	0.940902328792865\\
                    0.602714044568395	0.97054639146133\\
                    0.642492004094836	0.985121484428394\\
                    0.681903960975093	0.991885909730809\\
                    0.720972157381174	0.996528353607087\\
                    0.758918485506697	0.998527626623243\\
                    0.79438555077758	0.999311147038025\\
                    0.828051058093828	0.999637626198076\\
                    0.858749558957029	0.999767202475412\\
                    0.887416802421002	0.999832461450082\\
                    0.91175912480234	0.999872156456342\\
                    0.933064041114323	0.999904201270797\\
                    0.950664391873707	0.999930022351948\\
                    0.964995939253303	0.99995089103947\\
                    0.976078215946366	0.999966854698\\
                    0.984291620460386	0.99997852918788\\
                    0.990238139951513	0.999986871686049\\
                    0.994189287271368	0.999992318460874\\
                    0.996756129429708	0.999995799753463\\
                    };
                    \addlegendentry{$(64,30)$ eBCH, LC-SOSD}
                    
                    \addplot [color=mycolor3]
                      table[row sep=crcr]{%
                    0	0\\
                    0.0673261884696666	0.077566672811511\\
                    0.0839655667303307	0.101236071536972\\
                    0.105575820107303	0.131252258210212\\
                    0.129449177385451	0.166270887156782\\
                    0.162146676122568	0.220794911368604\\
                    0.197416054980562	0.279274427770264\\
                    0.24037034026258	0.352193014265149\\
                    0.291043790704039	0.438301230649729\\
                    0.348965503223697	0.541338164213025\\
                    0.381221472761545	0.594387774852145\\
                    0.415315571378787	0.647398198052565\\
                    0.448810103800074	0.697185298087531\\
                    0.485017522586853	0.747270527748984\\
                    0.523809212791884	0.796815836127707\\
                    0.561966410938583	0.839584534641903\\
                    0.601569102815201	0.87809853256601\\
                    0.644769388951565	0.912701677175094\\
                    0.681603740574896	0.937106922885181\\
                    0.719341222591274	0.957010133460773\\
                    0.758376534316889	0.973317276506199\\
                    0.794259391331887	0.983597172447002\\
                    0.827949193719952	0.990683633018704\\
                    0.859733202366459	0.99525805242422\\
                    0.887686948397641	0.997641023506777\\
                    0.911485965153349	0.998934090954895\\
                    0.933275231537332	0.999577150356605\\
                    0.950289198201806	0.99984454606969\\
                    0.964719661892858	0.999950206817314\\
                    0.976185291989449	0.999987074721857\\
                    0.984422793371353	0.999996965446282\\
                    0.990204959275385	0.999999396757221\\
                    0.994117991811645	0.999999898512943\\
                    0.996728852873088	0.999999986969643\\
                    };
                    \addlegendentry{$(8,4)$ eBCH, SOSD \cite{fossorier1998soft}}
                    
                    \addplot [color=mycolor3, dashed]
                      table[row sep=crcr]{%
                    0	0\\
                    0.0697399745254718	0.0787201558356745\\
                    0.0839851478036232	0.0941663576310413\\
                    0.104618025369135	0.123533588703803\\
                    0.131497886492896	0.15967734626989\\
                    0.161153966967641	0.204050924359487\\
                    0.198052213908218	0.25617406651862\\
                    0.242027830976901	0.327511442168367\\
                    0.292260150868156	0.406356335813253\\
                    0.347628251020232	0.504604577088772\\
                    0.381220279935628	0.56109241157025\\
                    0.414407603576663	0.612370709975892\\
                    0.448809466539194	0.662594103074211\\
                    0.489347000973902	0.721666172212204\\
                    0.522921978460495	0.763956654224607\\
                    0.565060875922603	0.811953431211357\\
                    0.602527990503557	0.849850174567065\\
                    0.643560372137996	0.885867602105706\\
                    0.683761504941115	0.915240510312701\\
                    0.721403472326407	0.940506533758124\\
                    0.759176980151568	0.960503878150142\\
                    0.795041946436179	0.975575271236988\\
                    0.827794818327293	0.985540831711191\\
                    0.859244489517673	0.992147419709605\\
                    0.888098844938148	0.99617228759132\\
                    0.911610839956779	0.998203309559574\\
                    0.932749334768024	0.999251109141464\\
                    0.950830259258576	0.999734824314441\\
                    0.964887314952481	0.999913440028169\\
                    0.975973760372516	0.999976084506155\\
                    0.98425261432352	0.999994393632315\\
                    0.990294440809988	0.999998903273674\\
                    0.994132325227298	0.999999765973913\\
                    0.996818349067945	0.999999952211744\\
                    };
                    \addlegendentry{$(8,4)$ eBCH, LC-SOSD}
                    
                    \addplot [color=mycolor4, dashed]
                      table[row sep=crcr]{%
                    0	0\\
                    1	1\\
                    };
                    \addlegendentry{$\mathcal{I}_{M}(\sigma_{\ell}^2) = \mathcal{I}_{M}(\sigma_{\delta}^2)$}
                    
                    \end{axis}
                    \end{tikzpicture}%
                    \vspace{-0.61em}
                \caption{MI Transform}     
                \vspace{-0.61em}
                \label{Fig::MI-trans}

             \end{subfigure}
             \hspace{-0.5em}
             \begin{subfigure}[b]{0.49\columnwidth}
                \centering
                \definecolor{mycolor1}{rgb}{0.00000,0.44706,0.74118}%
                \definecolor{mycolor2}{rgb}{0.49412,0.18431,0.55686}%
                \begin{tikzpicture}
                
                \begin{axis}[%
                    width=1.39in,
                    height=1.5in,
                    at={(0.788in,0.49in)},
                    scale only axis,
                    xmin=0.0,
                    xmax=1.0,
                    xlabel style={at={(0.5,2ex)},font=\color{white!15!black},font = \scriptsize},
                    xlabel={$\mathcal{I}_{M}(\sigma_{\ell}^2)$},
                    ymode = log,
                    ymin=0.1,
                    ymax=10000,
                    ylabel style={at={(4ex,0.5)},font=\color{white!15!black},font = \scriptsize},
                    ylabel={Average number of TEPs},
                    axis background/.style={fill=white},
                    tick label style={font=\tiny},
                    xmajorgrids,
                    ymajorgrids,
                    legend style={at={(0,0)}, anchor=south west, legend cell align=left, align=left, draw=white!15!black,font = \tiny,row sep=-3pt}
                ]
                \addplot [color=mycolor1]
                  table[row sep=crcr]{%
                0	4526\\
                1	4526\\
                };
                \addlegendentry{ $(64,30)$ eBCH, SOSD \cite{fossorier1998soft}}
                
                \addplot [color=mycolor1, dashed]
                  table[row sep=crcr]{%
                0	2100\\
                0.0675311228499975	2100.002\\
                0.0768134761891756	2103.658\\
                0.0874474217054265	2107.192\\
                0.0939593249247652	2103.584\\
                0.105228933274204	2106.679\\
                0.118928902463078	2100.947\\
                0.131316252702051	2009.541\\
                0.146591106391009	2103.946\\
                0.161227683013558	2104.445\\
                0.178768702145979	2102.412\\
                0.198053200876918	2152.933\\
                0.216944417697187	2247.552\\
                0.23877261818656	2224.065\\
                0.265027300718503	2251.734\\
                0.291542585172576	2320.659\\
                0.318482271661382	2235.951\\
                0.34941093878876	2219.775\\
                0.380841696669195	2182.948\\
                0.414600029340619	2065.538\\
                0.44894920925153	1789.274\\
                0.485253981578301	1549.782\\
                0.523271883159055	1200.224\\
                0.561080520928731	861.753\\
                0.602875223632171	591.475\\
                0.639414125821778	389.59\\
                0.680867731886733	228.464\\
                0.720785671690079	144.652\\
                0.759675640413832	78.062\\
                0.794308433689336	49.783\\
                0.828737736485642	39.966\\
                0.858848601992943	33.351\\
                0.904841217975947	31\\
                0.975881499145496	31\\
                0.984145202213394	31\\
                0.990208966478473	31\\
                0.994145148404032	31\\
                0.996758645860931	31\\
                };
                \addlegendentry{$(64,30)$ eBCH, LC-SOSD}
                
                \addplot [color=mycolor2]
                  table[row sep=crcr]{%
                0	11\\
                1	11\\
                };
                \addlegendentry{$(8,4)$ eBCH, SOSD \cite{fossorier1998soft}}
                
                \addplot [color=mycolor2, dashed]
                  table[row sep=crcr]{%
                0	6.512\\
                0.0694360543111776	6.512\\
                0.0762267459999614	6.4035\\
                0.0867903893443843	6.463\\
                0.0955155633432948	6.415\\
                0.107932656294413	6.404\\
                0.118409605833851	6.472\\
                0.131210422647974	6.4425\\
                0.146020519044447	6.513\\
                0.159193147705548	6.4355\\
                0.178076576239788	6.496\\
                0.194071348611804	6.5015\\
                0.219156033067978	6.461\\
                0.24567220371452	6.56\\
                0.262623325936462	6.42\\
                0.293737326695716	6.4085\\
                0.323747462631784	6.305\\
                0.343981960511429	6.3405\\
                0.376449347617887	6.367\\
                0.41170223383337	6.4235\\
                0.446773755609398	6.373\\
                0.485308280309162	6.311\\
                0.521913213403466	6.2305\\
                0.564077981178089	6.105\\
                0.597918100666372	6.045\\
                0.643386103175597	5.8645\\
                0.680165096809856	5.77\\
                0.718040013989249	5.6425\\
                0.756156577254768	5.513\\
                0.795582131477032	5.3315\\
                0.824267021103255	5.216\\
                0.858994132849569	5.1195\\
                0.888098844938148	5.1548\\
                0.911610839956779	5.0792\\
                0.932749334768024	5.0468\\
                0.950830259258576	5.018\\
                0.964887314952481	5.0108\\
                0.975973760372516	5.0012\\
                0.98425261432352	5\\
                0.990294440809988	5\\
                0.994132325227298	5\\
                0.996818349067945	5\\
                };
                \addlegendentry{$(8,4)$ eBCH, LC-SOSD}
                
                \end{axis}
                \end{tikzpicture}%
                \vspace{-0.61em}
                 \caption{Number of TEPs}
                 \vspace{-0.61em}
                 \label{Fig::MI-TEP}
             \end{subfigure}
             
             \caption{Comparisons between LC-SOSD and original SOSD}
             \vspace{-0.51em}
             \label{fig: graphs}
        \end{figure}

        We compare the outputs of LC-SOSD and the original SOSD by comparing their MI transform features. Assuming that the input LLR $\bm{\ell}$ follows a Gaussian distribution with variance $\sigma_{\ell}^2$, the input MI is given by \cite[Eq. (14)]{ten2001convergence}
        \begin{equation} \label{LowSISO::MI}
            \mathcal{I}_{M}(\sigma_{\ell}^2) =   \frac{1}{\sqrt{2\pi\sigma_{\ell}^2}}\int_{-\infty}^{+\infty}e^{-\frac{(\xi-\sigma_{\ell}^2/2)^2}{2\sigma_{\ell}^2}}\log_2(1+e^{-\xi})d\xi.
        \end{equation}
        MI reflects the divergence between the transmitted symbol and the estimated symbol evaluated from LLR \cite{ten2001convergence}.
        Assume that the $\bm{\delta}$ is also Gaussian and has the variance $\sigma_{\delta}^2$. Then, $\bm{\delta}$ introduces the output MI $\mathcal{I}_{M}(\sigma_{\delta}^2)$. The Gaussian assumption of $\delta_{i}$ has been widely applied and validated in the EXIT-chart analyses \cite{ten2001convergence}. In Fig. \ref{Fig::MI-trans}, we compare $\mathcal{I}_{M}(\sigma_{\delta}^2)$ as a function of $\mathcal{I}_{M}(\sigma_{\ell}^2)$. In LC-SOSD, we set $\lambda_s = 0.99$, and $\lambda_p$ is set according to \cite{yue2021probability}. The $(8,4,4)$ eBCH code and $(64,32,14)$ are decoded with order-2 and order-3 decoders, respectively. As shown, LC-SOSD has a very similar MI transform feature to the original SOSD. From Fig. \ref{Fig::MI-TEP}, when $\mathcal{I}_{M}(\sigma_{\ell}^2)$ is larger than 0.5 (i.e., moderate-to-high SNRs), LC-SOSD can significantly reduce the number of required TEPs, resulting in a considerably lower complexity. For example, in decoding $(64,30)$ eBCH, LC-SOSD requires 31 TEPs, compared to 4526 TEPs required by the original SOSD.

	\vspace{-0.2em} 
	\section{NOMA JD Receiver with LC-SOSD} \label{Sec::Receiver}
	\vspace{-0.2em}
    In this section, we elaborate on the details of the iterative JD receiver, shown in Fig. \ref{Fig::Structure}. 
    
    \vspace{-0.3em}
    \subsection{Parallel Interference Canceller}
    \vspace{-0.3em}
     We apply PIC to perform MUD for the considered short block-length regime, as PIC has been shown to nearly approach MAC capacity in MIMO and MIMO-NOMA systems in the large block-length scenarios \cite{liu2019capacity,kosasih2021bayesian}. Taking the procedure for user $u$ as an example, the priori information $\bm{\epsilon}^{(u)}(t-1)$ is fed to PIC at the beginning of iteration $t$, $t>1$. We initialize $\bm{\epsilon}^{(u)}(0) = 0$ for the first iteration. For the $i$-th transmitted symbol of user $u$, $x_{i}^{(u)}$, PIC estimates its mean and variance, respectively, as \cite{kosasih2021bayesian}
     \begin{equation}     \label{equ::Receiver::Primean}
         \mu_{i}^{(u)} = \tanh\left(\frac{\epsilon_{i}^{(u)}(t-1)}{2}\right) \ \ \text{and}  \ \  \upsilon_{i}^{(u)} =  1 - \left(\mu_{i}^{(u)}\right)^2.
     \end{equation}
     Next, PIC estimates the extrinsic LLR of the interference-removed symbol $\bar{x}_i^{(u)}$ according to \cite{wang2019near,kosasih2021bayesian}
    \begin{equation}  \label{equ::Receiver::PICLLR}
         \ell_{i}^{(u)}(t)  = \frac{2\mathrm{Re}\left(\frac{1}{h^{(u)}}\left(r_i - \sum_{j\neq u}h^{(j)}\mu_{i}^{(u)}\right)\right)}{\sum_{j\neq u}\left(\mathrm{Re}\left(\frac{h^{(j)}}{h^{(u)}}\right)\right)^2 \upsilon_{i}^{(u)} + \frac{\sigma^2}{2}}.
    \end{equation}
    We note that (\ref{equ::Receiver::PICLLR}) is obtained by assuming the interference as a Gaussian variable. This assumption, however, may not hold true when 1) user number is small and 2) the receiving power of users is significantly different. Despite this, we  still apply (\ref{equ::Receiver::PICLLR}) because of its computational efficiency.
    
    A decision statistics combiner (DSC) is usually implemented with PIC to smooth the convergence behavior\cite{marinkovic2001space,kosasih2021bayesian}. DSC generally combines the extrinsic LLRs of adjacent iterations with a parameter, $\beta$ ($0\leq\beta\leq1$), i.e.,
    \begin{equation}  \label{equ::Receiver::DSC}
    \begin{split}
         \ell_{i}^{(u)}(t) \!\leftarrow \! \tanh^{\!-1}\!\!\left(\!\beta\tanh\!\!\left(\!\!\frac{\ell_{i}^{(u)}(t)}{2}\!\!\right)\!\!+\!\!(1\!-\!\beta)\!\tanh\!\!\left(\!\!\frac{\ell_{i}^{(u)}(t-1)}{2}\!\!\right)\!\right)\!\!, 
    \end{split}
    \end{equation}  
    where $\tanh^{-1}(x)$ is the inverse of $\tanh(x)$. $\beta$ can be chosen constantly or adaptively according to MSE of $\bar{x}_i^{(u)}$. we select $\beta = 0.5$ in the proposed receiver for simplicity. Finally, the vector $\bm{\ell}^{(u)}(t) = [\ell^{(u)}(t)]_1^n$ is output by PIC.
    \vspace{-0.3em} 
    \subsection{Decoding Switch} \label{Sec::DS}
    \vspace{-0.3em}
    
       As shown by (\ref{equ::LowSISO::transMAP}), it is critical to correctly find the optimal codeword $\hat{\mathbf{c}}_{\mathrm{op}}$ for computing the extrinsic LLR $\delta_{i}$. Otherwise, both the sign and magnitude of $\delta_{i}$ might be incorrect.
       The probability that SOSD finds the incorrect $\hat{\mathbf{c}}_{\mathrm{op}}$ can be represented as $\mathrm{P}_{\mathrm{e}} \leq \mathrm{P}_{\mathrm{list}} + \mathrm{P}_{\mathrm{ML}}$ \cite{Fossorier1995OSD},
       where $\mathrm{P}_{\mathrm{list}}$ represents the probability that the transmitted codeword is not included in the list of codeword estimates generated by SOSD, and $\mathrm{P}_{\mathrm{ML}}$ is the ML error probability of $\mathcal{C}(n,k)$. $\mathrm{P}_{\mathrm{ML}}$ is determined by the code structure and its minimum distance $d_{\mathrm{H}}$, or obtained by available theoretical bounds of block codes, e.g., the tangential sphere bound (TSB)  \cite{poltyrev1994bounds}. $\mathrm{P}_{\mathrm{list}}$ is determined by the decoding order and the reliability of the input signal \cite{dhakal2016error}. In general, $\mathrm{P}_{\mathrm{list}}$ is large when input signal has a low reliability. PIC usually cannot properly remove the interference at early iterations due to inaccuracy of prior information $\bm{\epsilon}^{(u)}$. Consequently, the decoder output, $\bm{\delta}^{(u)}$, may result in an even worse BER compared to the decoder input, $\bm{\ell}^{(u)}$.
       
       Motivated by this, we design a DS to determine the engagement of SOSD in JD iterations. Specifically, when PIC fails to cancel the MAI properly and produces low-quality LLRs, the DS is set to``off''. When the LLR quality at PIC improves after a few iterations of MAI cancellation, the DS is turned ``on'' and decoding begins. Due to the space limit, we introduce a ``simple DS''. That is, the receiver performs $N_u$ iterations without decoding, and then turns on DS for subsequent iterations. We will further demonstrate the effeteness of ``Simple DS'' via simulations in Section \ref{Sec::Simulation}. One can further devise an adaptive DS based on the quality of PIC output.
       
    \vspace{-0.3em}
	\subsection{Decoding Combiner} \label{Sec::DC}
	\vspace{-0.3em}
	
        Even with DS enabled, it is possible that LC-SOSD may still produce unreliable posterior LLRs, because decoding errors cannot be completely avoided under finite SNRs according to the channel capacity theorem \cite{PPV2010l}. Hence, we can adaptively combine the decoder output, $\bm{\delta}^{(u)}$ with decoder input, $\bm{\ell}^{(u)}$, depending on the decoding quality. We define the decoding quality as $\gamma \triangleq \mathrm{P}_{\max}$ given by (\ref{equ::Define::Pmax}). This definition comes from that $\mathrm{P}_{\max}^{\mathrm{SP}}$ approximates the APP of $\hat{\mathbf{c}}_{\mathrm{op}}$ found by LC-SOSD. Thus, small $\mathrm{P}_{\max}^{\mathrm{SP}}$ indicates that $\hat{\mathbf{c}}_{\mathrm{op}}$ might not be the transmitted codeword. 
        
        The DC combines $\bm{\delta}^{(u)}$ and $\bm{\ell}^{(u)}$ according to the decoding quality $\gamma$ of each iteration. The combined LLR is given by 
        \begin{equation} \label{equ::Receiver::DC}
             \bm{\varphi}^{(u)}\!(t) \!=\! \tanh^{\!-1}\!\!\left(\!\gamma\tanh\!\left(\!\frac{\bm{\delta}^{(u)}(t)}{2}\!\right)\!+ \!(1\!-\!\gamma)\!\tanh\!\left(\!\frac{\bm{\ell}^{(u)}(t)}{2}\!\right)\!\right)\!\!, 
        \end{equation}  
        where $\tanh(\mathbf{x})$ and $\tanh^{-1}(\mathbf{x})$ are entry-wise operands of $\mathbf{x}$.
        
        We can see that both DS and DC are designed to improve the robustness of JD. DS is used priori to the decoder to avoid decoding low-quality inputs, leading to severe error propagation. DC is used after the decoder to reduce the impact of unreliable decoding.
        
        \subsection{Algorithm of the Iterative JD Receiver} 
        
        The decoding iteration is terminated when it exceeds a predetermined maximum number of iterations $t_{max}$, or the decoding results for all users are converged.  At iteration $t$, the decoding result of user $u$, denoted by $\hat{\mathbf{c}}^{(u)}(t)$, is obtained by the posterior LLR (addition of the extrinsic LLR and the prior LLR) of decoding, i.e.,
        \begin{equation}    \label{equ::Receiver::Ago::Est}
            \hat{c}_i^{(u)}(t) = 
            \begin{cases}
                0 \ \ \ \ \textup{if} \ \delta_{i}^{(u)}(t) + \ell_{i}^{(u)}(t) \geq 0, \\
                1 \ \ \ \ \textup{if} \ \delta_{i}^{(u)}(t) + \ell_{i}^{(u)}(t) < 0. \\
            \end{cases}
        \end{equation}
        The receiving iteration is stopped when $\hat{\mathbf{c}}^{(u)}(t) = \hat{\mathbf{c}}^{(u)}(t-1)$ holds for all users $1\leq u \leq N_u$. We summarize the JD receiver algorithm in Algorithm \ref{ago::Receiver}, where ``$\mathrm{DS} = 1$'' and  ``$\mathrm{DS} = 0$'' denote that DS is on and off, respectively.
        
        \vspace{-0.61em}
       \begin{algorithm} 
    	\caption{The Proposed JD Receiver}
    	\label{ago::Receiver}
    	\begin{algorithmic} [1]
    		\REQUIRE $\mathbf{r}$, $\mathbf{h}$, $t_{\max}$
    		\ENSURE The codeword estimates $\hat{\mathbf{c}}^{(1)},\ldots, \hat{\mathbf{c}}^{(N_u)}$ for all users
    		
    		\STATE Initialize $\bm{\epsilon}^{(u)}(0) \leftarrow 0$, $1\leq u \leq N_u$ , and $\mathrm{DS} \leftarrow 0$

    		\FOR{$t=1:t_{max}$}
    		\FOR{$u=1:N_u$ (in parallel)}
    		\STATE Compute $\bm{\mu}^{(u)}$ and $\bm{\upsilon}^{(u)}$ according to (\ref{equ::Receiver::Primean}).
    		\STATE Compute $\bm{\ell}^{(u)}(t)$ according to (\ref{equ::Receiver::PICLLR}).
    		\STATE \textbf{if} $t\geq 2$ \textbf{then} Update $\bm{\ell}^{(u)}(t)$ according to (\ref{equ::Receiver::DSC})
            \STATE \textbf{if} $t > N_u$ \textbf{then} $\mathrm{DS} \leftarrow 1$
    		\IF{$\mathrm{DS} = 1$}
    		\STATE Call Algorithm \ref{ago::LC-SISO}, input $\bm{\ell}^{(u)}(t)$ to acquire $\bm{\delta}^{(u)}(t)$
    		\STATE Obtain codeword estimates $\hat{\mathbf{c}}^{(u)}(t)$ by (\ref{equ::Receiver::Ago::Est})
			\STATE Obtain $\bm{\varphi}^{(u)}(t)$ according to (\ref{equ::Receiver::DC}) 
			\STATE $\bm{\epsilon}^{(u)}(t) \leftarrow \bm{\varphi}^{(u)}(t)$
    		\ELSE
    		\STATE $\bm{\epsilon}^{(u)}(t) \leftarrow \bm{\ell}^{(u)}(t)$
	     	\ENDIF
    		\ENDFOR		
    		\STATE \textbf{if} $\hat{\mathbf{c}}^{(u)}(t) = \hat{\mathbf{c}}^{(u)}(t-1) $ holds for all $u$ \textbf{then} \textbf{break}
    		\ENDFOR 
    		\RETURN $\hat{\mathbf{c}}^{(1)}(t),\ldots, \hat{\mathbf{c}}^{(N_u)}(t)$
    	\end{algorithmic}
        \end{algorithm}
        \vspace{-0.61em}
	
	\vspace{-0.2em}
	\section{Simulation and Comparisons} \label{Sec::Simulation}
	\vspace{-0.2em}
	In this section, we evaluate the BER and complexity of the proposed JD receiver. We assume that $\sum_{u=1}^{N_u}(\rho^{(u)})^2 = 1$. The ratio of the receiving power between any two adjacent users is assumed to be 4, i.e., $\frac{(\rho^{(u)})^2}{(\rho^{(u+1)})^2} = 4$ for $1\leq u \leq N_u-1$. The performance of our proposed scheme is compared to SIC decoding \cite{wang2004wireless}. SIC is embedded with the original (codeword-output) OSD to decode BCH codes for each user. That is, starting from the strongest undecoded user, SIC obtains its local optimal codeword by OSD, and then cancels the signal from the superposed received signal. Since we consider short BCH codes that approach NA, other existing approaches designed for moderate/long codes (e.g., LDPC and Polar codes) \cite{ping2004approaching,wang2019near,sharifi2015ldpc,balatsoukas2018design,ebada2020iterative,xiang2021iterative} are not compared in this paper. 
	\vspace{-0.3em}
	\subsection{Complexity Consideration}
	\vspace{-0.3em}
	PIC can efficiently perform the MAI cancellation with the complexity of $\mathcal{O}(n N_u)$ multiplications \cite{kosasih2021bayesian}. By considering the parallel architecture for each user, PIC is implemented with the complexity $C_{\mathrm{P}} = \mathcal{O}(n)$. On the other hand, the computational complexity of single OSD decoding can be expressed as \cite[Eq. (240)]{yue2021revisit} $C_{\mathrm{OSD}} =   \mathcal{O}(n) + \mathcal{O}(n\log n) + \mathcal{O}(n\min(k,n-k)^2)+ N_{a}\mathcal{O}(k+k(n-k))$, where $N_{a}$ is the number of TEPs re-encoded by OSD. Although LC-SOSD can significantly reduce the value of $N_{a}$, $C_{\mathrm{OSD}}$ is always higher than $\mathcal{O}(n\min(k,n-k)^2) + \mathcal{O}(k+k(n-k))$ by taking $N_a = 1$.
    
    Let $t_{\mathrm{off}}$ and $t_{\mathrm{on}}$ denote the average numbers of DS-off and DS-on iterations, respectively. Then, we represent the overall complexity of the proposed JD receiver as
    \begin{equation} \label{equ::Complexity}
        C_{\mathrm{JD}} = t_{\mathrm{off}}C_{\mathrm{P}} + t_{\mathrm{on}}(C_{\mathrm{P}} + C_{\mathrm{OSD}}) \approx t_{\mathrm{on}}(C_{\mathrm{P}} + C_{\mathrm{OSD}})
    \end{equation}
    because $C_{\mathrm{P}} \ll C_{\mathrm{OSD}}$. In contrast, the complexity of SIC approach can be approximately represented as $C_{\mathrm{SIC}} = N_u(\mathcal{O}(n) + C_{\mathrm{OSD}})$. Therefore, in the simulation, we mainly compare the complexity by comparing the number of decoding iterations, i.e., $N_u$ for SIC and $t_{\mathrm{on}}$ for the proposed receiver.
    
    Different from the sequential decoding behavior of SIC, the JD receiver decodes all users in parallel simultaneously. If $t_{\mathrm{on}} < N_u$, JD requires a lower number of decoding iterations than SIC to complete the decoding of all users, which results in a lower receiving latency. Additionally, the proposed LC-SOSD can reduce the complexity of each single decoding iteration, reducing the receiving latency even further.
    
    \subsection{Simulation Results}
    \subsubsection{DS and DC} We consider variants of the proposed JD receiver where DS and DC are (partially) removed. When DS is disabled, decoding starts at the first iteration, while when DC is disabled, $\bm{\epsilon}^{(u)}(t) \leftarrow \bm{\delta}^{(u)}(t)$ is directly fedback to PIC. We conduct simulations for the three-user NOMA over the fading channel with $(64,30,14)$ extended BCH (eBCH) code, decoded by order-3 decoder. As shown in Fig. \ref{Fig::DSDC-BER}, when DS and DC are removed, the proposed JD receiver has a BER performance degradation at high SNRs, because DS and DC can eliminate the effect of unreliable decoding. In terms of the complexity, the JD without DS and DC shows a high number of decoding iterations, up to 20. While employing both DS and DC, the JD receiver is more efficient than SIC and requires fewer decoding iterations.

            \begin{figure}
             \centering
             \vspace{-0.25em}
             \hspace{-0.81em}
             \begin{subfigure}[b]{0.49\columnwidth}
                 \centering
                \begin{tikzpicture}
                \begin{axis}[%
                width=1.35in,
                height=1.5in,
                at={(0.785in,0.587in)},
                scale only axis,
                xmin=0,
                xmax=20,
                xlabel style={at={(0.5,2ex)},font=\color{white!15!black},font=\scriptsize},
                xlabel={SNR (dB)},
                ymode=log,
                ymin=5e-03,
                ymax=1,
                yminorticks=true,
                ylabel style={at={(3ex,0.5)},font=\color{white!15!black},font=\scriptsize},
                ylabel={Average BER},
                axis background/.style={fill=white},
                tick label style={font=\tiny},
                xmajorgrids,
                ymajorgrids,
                yminorgrids,
                minor grid style={dotted},
                major grid style={dotted,black},
                legend style={at={(0,0)}, anchor=south west, legend cell align=left, align=left, draw=white!15!black,font = \tiny,row sep=-3pt}
                ]
                \addplot [color=blue, mark=square,mark size=1pt, mark options={solid, blue}]
                  table[row sep=crcr]{%
                    0	0.315314180107527\\
                    2	0.285583941605839\\
                    4	0.238126899696049\\
                    6	0.187624601275917\\
                    8	0.155751992031873\\
                    10	0.116728855721393\\
                    12	0.0883239171374764\\
                    14	0.0629776749798874\\
                    16	0.0421841252699784\\
                    18	0.0288834503140007\\
                    20	0.0228042549826004\\
                };
                \addlegendentry{SIC decoding}
                
                \addplot [color=red, mark=square,mark size=1pt, mark options={solid, red}]
                  table[row sep=crcr]{%
                    0	0.3\\
                    2	0.268\\
                    4	0.216\\
                    6	0.177\\
                    8	0.129\\
                    10	0.102\\
                    12	0.0714\\
                    14	0.051\\
                    16	0.0368\\
                    18	0.0242\\
                    20	0.0176\\
                };
                \addlegendentry{DS removed, DC removed}
                
                \addplot [color=red, mark=triangle, mark options={solid, red}]
                  table[row sep=crcr]{%
                    0	0.298690476190476\\
                    2	0.262083333333333\\
                    4	0.218229166666667\\
                    6	0.167501335470085\\
                    8	0.142001589464124\\
                    10	0.0915897773872291\\
                    12	0.0656145309882747\\
                    14	0.043886899747262\\
                    16	0.0286630036630037\\
                    18	0.0195132398753894\\
                    20	0.011948459126711\\
                };
                \addlegendentry{DS removed, DC employed}
                
                \addplot [color=red, mark=*,mark size = 1pt, mark options={solid, red}]
                  table[row sep=crcr]{%
                    0	0.301402243589744\\
                    2	0.255225298588491\\
                    4	0.215335169880624\\
                    6	0.172981194690266\\
                    8	0.128403804050356\\
                    10	0.0945309351068118\\
                    12	0.0657502803476311\\
                    14	0.0389341712294674\\
                    16	0.0266731948228883\\
                    18	0.0169318755863463\\
                    20	0.0105375\\
                };
                \addlegendentry{DS and DC employed}
                \end{axis}
                \end{tikzpicture}%
                \vspace{-0.61em}
                \caption{BER performance}  
                \vspace{-0.61em}
                \label{Fig::DSDC-BER}

             \end{subfigure}
             \hspace{-0.4em}
             \begin{subfigure}[b]{0.49\columnwidth}
                \centering
                \begin{tikzpicture}
                \begin{axis}[%
                width=1.35in,
                height=1.5in,
                at={(0.642in,0.505in)},
                scale only axis,
                xmin=0,
                xmax=20,
                xlabel style={at={(0.5,2ex)},font=\color{white!15!black},font=\scriptsize},
                xlabel={SNR (dB)},
                ymode=log,
                ymin=1,
                ymax=100,
                ylabel style={at={(4ex,0.5)},font=\color{white!15!black},font=\scriptsize},
                ylabel={Average number of  decoding iterations},
                axis background/.style={fill=white},
                tick label style={font=\tiny},
                xmajorgrids,
                ymajorgrids,
                yminorgrids,
                minor grid style={dotted},
                major grid style={dotted,black},
                legend style={at={(1,1)}, anchor=north east, legend cell align=left, align=left, draw=white!15!black,font = \tiny,row sep=-3pt}
                ]
                \addplot [color=blue, line width = 1pt]
                  table[row sep=crcr]{%
                0	3\\
                20	3\\
                };
                \addlegendentry{SIC decoding}
                
                \addplot [color=red, mark=square,mark size = 1pt,  mark options={solid, red}]
                  table[row sep=crcr]{%
                    0	19.8793103448276\\
                    2	23.105083446148\\
                    4	22.6104294679495\\
                    6	22.0690772029056\\
                    8	20.2840240617923\\
                    10	16.5791858226844\\
                    12	12.9782757946356\\
                    14	10.8844418223666\\
                    16	8.75050100848767\\
                    18	7.25534064997604\\
                    20	6.50284495021337\\
                };
                \addlegendentry{DS removed, DC removed}

                \addplot [color=red, mark=triangle, mark options={solid, red}]
                  table[row sep=crcr]{%
                    0	4.78857142857143\\
                    2	6.43119047619048\\
                    4	7.36333945164735\\
                    6	7.87943887419317\\
                    8	7.41306198977106\\
                    10	6.95436780274495\\
                    12	6.02715168552883\\
                    14	5.11333121645653\\
                    16	4.38427548020836\\
                    18	4.01391091942764\\
                    20	3.72378986005965\\
                };
                \addlegendentry{DS removed, DC employed}
    
                \addplot [color=red, mark=*, mark size = 1pt, mark options={solid, red}]
                  table[row sep=crcr]{%
                    0	2.58\\
                    2	2.73\\
                    4	2.77\\
                    6	2.69\\
                    8	2.54\\
                    10	2.346\\
                    12	2.176\\
                    14	2.1\\
                    16	2.08\\
                    18	2.09\\
                    20	2.09\\
                };
                \addlegendentry{DS and DC employed}
                \end{axis}
                \end{tikzpicture}%
                \vspace{-0.61em}
                 \caption{Number of decoding iterations}
                 \vspace{-0.61em}
                 \label{Fig::DSDC-iter}
             \end{subfigure}
             
              \vspace{-0.11em}
             \caption{The BER performance and number of decoding iterations of different variants of the proposed JD receiver.}
            \vspace{-0.61em}
             \label{Fig::DSDC}
        \end{figure}
        
    \subsubsection{AWGN channel} 
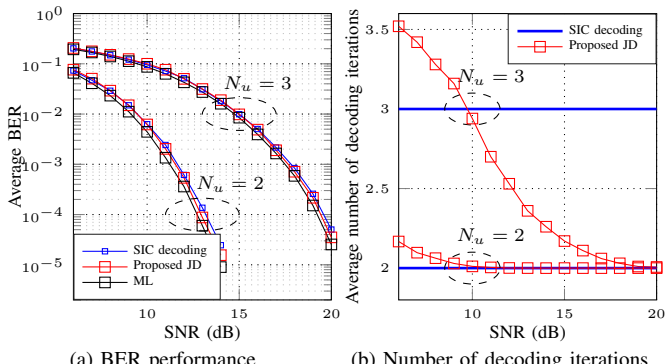
\begin{figure}
             \centering
             \hspace{-0.81em}
             \begin{subfigure}[b]{0.49\columnwidth}
                 \centering
                \begin{tikzpicture}
                
                \begin{axis}[%
                width=1.35in,
                height=1.5in,
                at={(0.785in,0.587in)},
                scale only axis,
                xmin=6,
                xmax=20,
                xlabel style={at={(0.5,2ex)},font=\color{white!15!black},font=\scriptsize},
                xlabel={SNR (dB)},
                ymode=log,
                ymin= 2e-06,
                ymax=1,
                yminorticks=true,
                ylabel style={at={(3ex,0.5)},font=\color{white!15!black},font=\scriptsize},
                ylabel={Average BER},
                axis background/.style={fill=white},
                tick label style={font=\tiny},
                xmajorgrids,
                ymajorgrids,
                yminorgrids,
                minor grid style={dotted},
                major grid style={dotted,black},
                legend style={at={(0,0)}, anchor=south west, legend cell align=left, align=left, draw=white!15!black,font = \tiny,row sep=-3pt}
                ]
                \addplot [color=blue, mark=square,mark size=1pt, mark options={solid, blue}]
                  table[row sep=crcr]{%
                6	0.202712926934717\\
                7	0.176077459855507\\
                8	0.147757976516525\\
                9	0.121153838944033\\
                10	0.0953981003586903\\
                11	0.0717084435734817\\
                12	0.0516678196931847\\
                13	0.0311097325504646\\
                14	0.0180917966879404\\
                15	0.0099368232077282\\
                16	0.00503663486546488\\
                17	0.0021580961872989\\
                18	0.00084134528526052\\
                19	0.00025855555555556\\
                20	0.00005033333333333\\
                };
                \addlegendentry{SIC decoding}
                
                \addplot [color=red, mark=square, mark options={solid, red}]
                  table[row sep=crcr]{%
                6	0.206543291473893\\
                7	0.177301486129502\\
                8	0.15093287514657\\
                9	0.124133793172163\\
                10	0.100132094228868\\
                11	0.0762934301015759\\
                12	0.0501423681509846\\
                13	0.0314823962192468\\
                14	0.0188907159342988\\
                15	0.00985793054013763\\
                16	0.00489751228862991\\
                17	0.00190288099481219\\
                18	0.00071222936875678\\
                19	0.000215833333333333\\
                20	0.000035166666666667\\
                };
                \addlegendentry{Proposed JD}
                
                \addplot [color=black, mark=square, mark options={solid, black}]
                  table[row sep=crcr]{%
                6	0.193918709277071\\
                7	0.168071634522972\\
                8	0.139926624715034\\
                9	0.1129020550139\\
                10	0.086922117127728\\
                11	0.0631417580604072\\
                12	0.0428000015300161\\
                13	0.027421599456271\\
                14	0.0160630259566201\\
                15	0.00847534007819163\\
                16	0.00393375071841024\\
                17	0.00164399637923607\\
                18	0.000588765321296145\\
                19	1.51388888888889e-04\\
                20	2.54166666666667e-05\\
                };
                \addlegendentry{ML}
                
                \addplot [color=blue, mark=square,mark size=1pt, mark options={solid, blue}]
                  table[row sep=crcr]{%
                6	0.0734085036410618\\
                7	0.0470088027565961\\
                8	0.0291072351865532\\
                9	0.0149490344583409\\
                10	0.0062829894356338\\
                11	0.00242320353259558\\
                12	0.00061552871801491\\
                13	0.0001365\\
                14	0.00002453\\
                };
                
                \addplot [color=red, mark=square, mark options={solid, red}]
                  table[row sep=crcr]{%
                6	0.076449482182174\\
                7	0.0506100194342475\\
                8	0.0291910574168639\\
                9	0.0149500071760034\\
                10	0.00620583373181228\\
                11	0.00202625\\
                12	0.00053750\\
                13	0.00008750\\
                14	1.555e-05\\
                15	0\\
                };

                \addplot [color=black, mark=square, mark options={solid, black}]
                  table[row sep=crcr]{%
                6	0.066039729501268\\
                7	0.0410142582225332\\
                8	0.0232634581989581\\
                9	0.0111634246803648\\
                10	0.0044275\\
                11	0.0013525\\
                12	0.00036125\\
                13	6e-05\\
                14	9.0025e-06\\
                15	0\\
                };
                
                \draw[dashed, color=black] 
                (axis cs:15,0.01) ellipse[ x radius = 20, y radius = 0.75];    
                \node[] at (axis cs: 16,0.04) {\scriptsize $N_u=3$};
                
                \draw[dashed, color=black] 
                (axis cs:13,0.0001) ellipse[ x radius = 20, y radius = 0.75];    
                \node[] at (axis cs: 14.5,0.0004) {\scriptsize $N_u=2$};
                
                \end{axis}
                \end{tikzpicture}%

                \vspace{-0.61em}
                \caption{BER performance}     
                \vspace{-0.61em}
                \label{Fig::8-4-BER-AWGN}

             \end{subfigure}
             \hspace{-0.3em}
             \begin{subfigure}[b]{0.49\columnwidth}
                \centering
                \begin{tikzpicture}
                
                \begin{axis}[%
                width=1.35in,
                height=1.5in,
                at={(0.642in,0.505in)},
                scale only axis,
                xmin=6,
                xmax=20,
                xlabel style={at={(0.5,2ex)},font=\color{white!15!black},font=\scriptsize},
                xlabel={SNR (dB)},
                ymin=1.8,
                ymax=3.6,
                ylabel style={at={(4ex,0.5)},font=\color{white!15!black},font=\scriptsize},
                ylabel={Average number of  decoding iterations},
                axis background/.style={fill=white},
                tick label style={font=\tiny},
                xmajorgrids,
                ymajorgrids,
                minor grid style={dotted},
                major grid style={dotted,black},
                legend style={at={(1,1)}, anchor=north east, legend cell align=left, align=left, draw=white!15!black,font = \tiny,row sep=-3pt}
                ]
                
                \addplot [color=blue, line width = 1pt]
                  table[row sep=crcr]{%
                6	3\\
                20	3\\
                };
                \addlegendentry{SIC decoding}

                \addplot [color=red, mark=square, mark options={solid, red}]
                  table[row sep=crcr]{%
                6	3.52\\
                7	3.42\\
                8	3.28\\
                9	3.16\\
                10	2.94\\
                11	2.7\\
                12	2.53\\
                13	2.36\\
                14	2.26\\
                15	2.17\\
                16	2.11\\
                17	2.06\\
                18	2.03\\
                19	2.01\\
                20	2.007\\
                };
                \addlegendentry{Proposed JD}

                \addplot [color=blue, line width = 1pt]
                  table[row sep=crcr]{%
                6	2\\
                20	2\\
                };
                
                \addplot [color=red, mark=square, mark options={solid, red}]
                  table[row sep=crcr]{%
                6	2.1671\\
                7	2.09696666666667\\
                8	2.06238\\
                9	2.02922\\
                10	2.01088\\
                11	2.00434\\
                12	2.00074\\
                13	2.00014\\
                14	1.99988\\
                15	2.00012\\
                16	2\\
                17	2\\
                18	2\\
                19	2\\
                20	2\\
                };
                
                \draw[dashed, color=black] 
                (axis cs:10,3) ellipse[ x radius = 15, y radius = 10];    
                \node[] at (axis cs: 11,3.2) {\scriptsize $N_u=3$};
                
                \draw[dashed, color=black] 
                (axis cs:10,2) ellipse[ x radius = 15, y radius = 10];    
                \node[] at (axis cs: 11,2.2) {\scriptsize $N_u=2$};
                
                \end{axis}
                \end{tikzpicture}%
                \vspace{-0.61em}
                 \caption{Number of decoding iterations}
                 \vspace{-0.61em}
                \label{Fig::8-4-iter-AWGN}
             \end{subfigure}
             \vspace{-0.11em}
             \caption{The comparisons of proposed JD and SIC decoding with $(8,4,4)$ eBCH code over AWGN channels}
             \vspace{-0.61em}
             \label{Fig::8-4-AWGN}
        \end{figure}

In the AWGN channel, each entry of $\mathbf{h}$ is set to $h^{(u)} = \rho^{(u)}$, $1\leq u \leq N_u$. We simulate the $(8,4,4)$ eBCH code. Despite this code is too short for practical systems, we can simulate its ML decoding BER as a performance benchmark, where ML results are obtained by exhausting the codebooks of all users. The average BER of order-2 decoding are compared in Fig. \ref{Fig::8-4-BER-AWGN}. It can be seen that the proposed JD receiver has a slightly better BER performance than SIC and approaches the ML performance. Fig. \ref{Fig::8-4-iter-AWGN} shows the complexity in terms of the number of decoding iterations. The proposed JD receiver requires fewer decoding iterations than SIC when $N_u = 3$, and has similar complexity to SIC when $N_u = 2$.
    
    \subsubsection{Fading channel} 
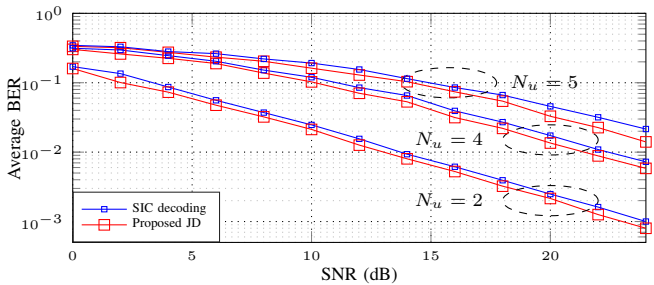
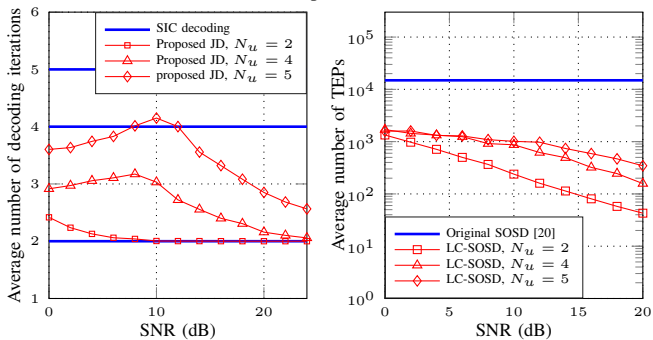
\begin{figure}
             \centering
             \begin{subfigure}[b]{1\columnwidth}
                 \centering
                \begin{tikzpicture}
                \begin{axis}[%
                width=3in,
                height=1.2in,
                at={(0.785in,0.587in)},
                scale only axis,
                xmin=0,
                xmax=24,
                xlabel style={at={(0.5,2ex)},font=\color{white!15!black},font=\scriptsize},
                xlabel={SNR (dB)},
                ymode=log,
                ymin=5e-04,
                ymax=1,
                yminorticks=true,
                ylabel style={at={(3ex,0.5)},font=\color{white!15!black},font=\scriptsize},
                ylabel={Average BER},
                axis background/.style={fill=white},
                tick label style={font=\tiny},
                xmajorgrids,
                ymajorgrids,
                yminorgrids,
                minor grid style={dotted},
                major grid style={dotted,black},
                legend style={at={(0,0)}, anchor=south west, legend cell align=left, align=left, draw=white!15!black,font = \tiny,row sep=-3pt}
                ]
                \addplot [color=blue, mark=square,mark size=1pt, mark options={solid, blue}]
                  table[row sep=crcr]{%
                    -4	0.291240706319703\\
                    -2	0.242066563467492\\
                    -4.82163733276644e-16	0.170646834061135\\
                    2	0.135326557093426\\
                    4	0.0867516629711752\\
                    6	0.05584345961401\\
                    8	0.0371381110583768\\
                    10	0.0246881907167667\\
                    12	0.0155349334128404\\
                    14	0.00933766427718041\\
                    16	0.006153125\\
                    18	0.00393125\\
                    20	0.002484375\\
                    22	0.001615625\\
                    24	0.001\\
                };
                \addlegendentry{SIC decoding}
                
                \addplot [color=red, mark=square, mark options={solid, red}]
                  table[row sep=crcr]{%
                    -4	0.39390625\\
                    -2	0.353230337078652\\
                    -4.82163733276644e-16	0.334574468085106\\
                    2	0.319387755102041\\
                    4	0.274021739130435\\
                    6	0.233488805970149\\
                    8	0.203814935064935\\
                    10	0.162048969072165\\
                    12	0.129218106995885\\
                    14	0.10351821192053\\
                    16	0.0739657210401891\\
                    18	0.0543185763888889\\
                    20	0.0328489634748272\\
                    22	0.0225681989449887\\
                    24	0.0140372775372775\\
                };
                \addlegendentry{Proposed JD}
                
                \addplot [color=blue, mark=square,mark size=1pt, mark options={solid, blue}]
                  table[row sep=crcr]{%
                    -4	0.384037990196078\\
                    -2	0.342927631578947\\
                    -4.82163733276644e-16	0.314\\
                    2	0.296401515151515\\
                    4	0.245676100628931\\
                    6	0.203938802083333\\
                    8	0.152010658914729\\
                    10	0.119871549079755\\
                    12	0.0850203804347826\\
                    14	0.0654264214046823\\
                    16	0.0392978162650602\\
                    18	0.0270465940525588\\
                    20	0.0173265272244356\\
                    22	0.0109000487736901\\
                    24	0.00724939781360015\\
                };
                
                \addplot [color=red, mark=square, mark options={solid, red}]
                  table[row sep=crcr]{%
                    -4	0.260382059800664\\
                    -2	0.209701742627346\\
                    -4.82163733276644e-16	0.160403885480573\\
                    2	0.100627413127413\\
                    4	0.0731308411214953\\
                    6	0.0474712295578437\\
                    8	0.0322012546277252\\
                    10	0.0212504640678865\\
                    12	0.0126242306025141\\
                    14	0.00801349847800987\\
                    16	0.00525625\\
                    18	0.003234375\\
                    20	0.00215\\
                    22	0.001259375\\
                    24	0.0008\\
                };
                
                \addplot [color=blue, mark=square,mark size=1pt, mark options={solid, blue}]
                  table[row sep=crcr]{%
                    -4	0.414967105263158\\
                    -2	0.386728395061728\\
                    -4.82163733276644e-16	0.345604395604396\\
                    2	0.329473684210526\\
                    4	0.281138392857143\\
                    6	0.262604166666667\\
                    8	0.220367132867133\\
                    10	0.192024539877301\\
                    12	0.155136138613861\\
                    14	0.11456043956044\\
                    16	0.0852861035422343\\
                    18	0.066427813163482\\
                    20	0.0456204379562044\\
                    22	0.0317687246963563\\
                    24	0.0215158293186511\\
                };
                
                \addplot [color=red, mark=square, mark options={solid, red}]
                  table[row sep=crcr]{%
                    -4	0.375901442307692\\
                    -2	0.343338815789474\\
                    -4.82163733276644e-16	0.302283653846154\\
                    2	0.259830298013245\\
                    4	0.22485632183908\\
                    6	0.189227764423077\\
                    8	0.137764084507042\\
                    10	0.10287140052356\\
                    12	0.0704948384201077\\
                    14	0.0531165081521739\\
                    16	0.0315016103059581\\
                    18	0.0220174451322454\\
                    20	0.0135294244728655\\
                    22	0.0088273849797023\\
                    24	0.00583806605888507\\
                };

                \draw[dashed, color=black] 
                (axis cs:20,0.002) ellipse[ x radius = 20, y radius = 0.5];    
                \node[] at (axis cs: 15.75,0.002) {\scriptsize $N_u=2$};
                
                \draw[dashed, color=black] 
                (axis cs:20,0.015) ellipse[ x radius = 20, y radius = 0.5];    
                \node[] at (axis cs: 15.75,0.015) {\scriptsize $N_u=4$};
                
                \draw[dashed, color=black] 
                (axis cs:15.75,0.1) ellipse[ x radius = 20, y radius = 0.5];    
                \node[] at (axis cs: 19.75,0.1) {\scriptsize $N_u=5$};
                
                \end{axis}
                \end{tikzpicture}%

                \vspace{-0.61em}
                \caption{BER performance}  
                \vspace{-0.151em}
                \label{Fig::64-16-BER-Fading}

             \end{subfigure}
             \hfill
             \hspace{-0.5em}
             \begin{subfigure}[b]{0.49\columnwidth}
                \centering
               \begin{tikzpicture}
                \begin{axis}[%
                width=1.35in,
                height=1.5in,
                at={(0.642in,0.505in)},
                scale only axis,
                xmin=0,
                xmax=24,
                xlabel style={at={(0.5,2ex)},font=\color{white!15!black},font=\scriptsize},
                xlabel={SNR (dB)},
                ymin=1,
                ymax=6,
                ytick distance=1,
                ylabel style={at={(5ex,0.5)},font=\color{white!15!black},font=\scriptsize},
                ylabel={Average number of decoding iterations},
                axis background/.style={fill=white},
                tick label style={font=\tiny},
                xmajorgrids,
                ymajorgrids,
                minor grid style={dotted},
                major grid style={dotted,black},
                legend style={at={(1,1)}, anchor=north east, legend cell align=left, align=left, draw=white!15!black,font = \tiny,row sep=-3pt}
                ]
                \addplot [color=blue, line width = 1pt]
                  table[row sep=crcr]{%
                0	2\\
                24	2\\
                };
                \addlegendentry{SIC decoding}
                
                \addplot [color=red, mark=square,mark size=1pt, mark options={solid, red}]
                  table[row sep=crcr]{%
                    -0	2.41717791411043\\
                    2	2.23603603603604\\
                    4	2.12803738317757\\
                    6	2.06056935190793\\
                    8	2.03990127519539\\
                    10	2.00371254309202\\
                    12	2.00202282907094\\
                    14	2.00199433189881\\
                    16	2.0011\\
                    18	2.0005\\
                    20	2.0001\\
                    22	2.0001\\
                    24	2\\
                };
                \addlegendentry{Proposed JD, $N_u=2$}
                
                \addplot [color=blue, line width = 1pt,forget plot]
                  table[row sep=crcr]{%
                0	4\\
                24	4\\
                };

                \addplot [color=red, mark=triangle, mark options={solid, red}]
                  table[row sep=crcr]{%
                    0	2.91538461538462\\
                    2	2.97218543046358\\
                    4	3.0563218390805\\
                    6	3.10673076923077\\
                    8	3.17253521126761\\
                    10	3.03141361256545\\
                    12	2.72244165170557\\
                    14	2.55603260869565\\
                    16	2.40001610305958\\
                    18	2.30725942599887\\
                    20	2.15865883166263\\
                    22	2.10239061795219\\
                    24	2.05768943356748\\
                };
                \addlegendentry{Proposed JD, $N_u=4$}

                \addplot [color=blue, line width = 1pt,forget plot]
                  table[row sep=crcr]{%
                0	5\\
                24	5\\
                };
                
                \addplot [color=red, mark=diamond, mark options={solid, red}]
                  table[row sep=crcr]{%
                    0	3.6031914893617\\
                    2	3.63469387755102\\
                    4	3.74347826086957\\
                    6	3.82985074626866\\
                    8	4.01428571428571\\
                    10	4.15051546391753\\
                    12	4.00119341563786\\
                    14	3.55629139072848\\
                    16	3.32033096926714\\
                    18	3.08229166666667\\
                    20	2.85143139190523\\
                    22	2.68432554634514\\
                    24	2.56459836459836\\
                };
                \addlegendentry{proposed JD, $N_u=5$}
                \end{axis}
                \end{tikzpicture}%
                \vspace{-0.61em}
                 \caption{Number of decoding iterations}
                 \vspace{-0.31em}
                \label{Fig::64-16-iter-Fading}
             \end{subfigure}
             \hspace{-0.5em}
             \begin{subfigure}[b]{0.49\columnwidth}
                \centering
                \begin{tikzpicture}
                \begin{axis}[%
                width=1.35in,
                height=1.5in,
                at={(0.642in,0.505in)},
                scale only axis,
                xmin=0,
                xmax=20,
                xlabel style={at={(0.5,2ex)},font=\color{white!15!black},font=\scriptsize},
                xlabel={SNR (dB)},
                ymode=log,
                ymin=1,
                ymax=3*10^5,
                ylabel style={at={(4ex,0.5)},font=\color{white!15!black},font=\scriptsize},
                ylabel={Average number of TEPs},
                axis background/.style={fill=white},
                tick label style={font=\tiny},
                xmajorgrids,
                ymajorgrids,
                minor grid style={dotted},
                major grid style={dotted,black},
                legend style={at={(0,0)}, anchor=south west, legend cell align=left, align=left, draw=white!15!black,font = \tiny,row sep=-3pt}
                ]
                \addplot [color=blue, line width = 1pt]
                  table[row sep=crcr]{%
                0	14893\\
                20	14893\\
                };
                \addlegendentry{Original SOSD \cite{fossorier1998soft}}
                
                \addplot [color=red, mark=square,mark size=1.5pt, mark options={solid, red}]
                  table[row sep=crcr]{%
                    -4	1819.71031284607\\
                    -2	1566.65658432274\\
                    -4.82163733276644e-16	1339.49727861756\\
                    2	974.890679710324\\
                    4	712.73853888518\\
                    6	499.955293495975\\
                    8	366.564825221775\\
                    10	238.933211318204\\
                    12	160.106254085219\\
                    14	113.7126355171\\
                    16	80.5780928571419\\
                    18	57.7541523809528\\
                    20	42.8508166666673\\
                    22	34.6938833333336\\
                    24	25.4064666666662\\
                };
                \addlegendentry{LC-SOSD, $N_u = 2$}
                
                \addplot [color=red, mark=triangle, mark options={solid, red}]
                  table[row sep=crcr]{%
                    -4	1754.9721998054\\
                    -2	1827.81539035292\\
                    -4.82163733276644e-16	1699.75883179321\\
                    2	1439.16528144312\\
                    4	1329.37737454977\\
                    6	1244.76185224048\\
                    8	916.566552122076\\
                    10	871.089685616661\\
                    12	619.374326677018\\
                    14	492.850122182128\\
                    16	321.61189109618\\
                    18	244.909858383191\\
                    20	157.870950081802\\
                    22	107.901176335817\\
                    24	77.4486525724019\\
                };
                \addlegendentry{LC-SOSD, $N_u = 4$}
                
                \addplot [color=red, mark=diamond, mark options={solid, red}]
                  table[row sep=crcr]{%
                    -4	1824.30249185999\\
                    -2	1697.63474092319\\
                    -4.82163733276644e-16	1585.9525268799\\
                    2	1598.91359552521\\
                    4	1315.60677239186\\
                    6	1285.72567284137\\
                    8	1091.39757107012\\
                    10	1012.83697930369\\
                    12	973.66233954938\\
                    14	739.544266132525\\
                    16	590.639859275478\\
                    18	468.543626835075\\
                    20	345.554080092572\\
                    22	245.142690849888\\
                    24	174.083298300712\\
                };
                \addlegendentry{LC-SOSD, $N_u = 5$}
                \end{axis}
                \end{tikzpicture}%
                \vspace{-0.61em}
                 \caption{Number of TEPs}
                 \vspace{-0.31em}
                \label{Fig::64-16-TEP-Fading}
             \end{subfigure}
              \vspace{-0.2em}
             \caption{The comparisons of proposed JD and SIC decoding with $(64,16,24)$ eBCH code over fading channels.}
            \vspace{-0.61em}
             \label{Fig::64-16-Fading}
        \end{figure}

Low-rate codes are commonly used in fading channels to prevent severe MAI\cite{ping2004approaching}. We simulate the transmission with low-rate $(64,16,24)$ eBCH code with order-6 decoding, as depicted in Fig \ref{Fig::64-16-Fading}. As shown, the proposed JD receiver reaches a lower BER than the SIC in the fading channel, and over 2 dB gain of BER performance is observed when $N_u = 5$. From Fig. \ref{Fig::64-16-iter-Fading}, the JD receiver significantly reduces the number of decoding iterations. When $N_u = 5$, for example, it performs less than 3 decoding iterations compared to 5 of SIC. Furthermore, numbers of re-encoded TEPs in single decoding are summarized in Fig. \ref{Fig::64-16-TEP-Fading}. Originally, 14893 TEPs are required in order-$6$ SOSD for decoding $(64,16,24)$ eBCH; however, the applied LC-SOSD only requires a few hundreds TEPs at moderate-to-high SNRs.

\vspace{-0.3em}
\section{Conclusion} \label{sec::Conclusion}
\vspace{-0.3em}

In this paper, we designed an efficient joint decoding (JD) receiver for power-domain NOMA systems for short block length codes. We proposed a low-complexity soft-output OSD (LC-SOSD) to reduce the complexity of the original SOSD. Using an early stopping condition, the decoding processes were stopped when the success probabilities of codewords satisfy a threshold. Then, for NOMA systems, an efficient iterative JD receiver was devised by combining parallel interference cancellation and the proposed LC-SOSD. Two novel techniques: decoding switch and decoding combiner, were introduced to accelerate the convergence. Several simulations with short BCH codes show that the proposed JD achieves better bit-error-rate performance than SIC over the AWGN and fading channel, while exhibiting a lower receiving complexity.






%
	\vspace{-0.3em}
\bibliographystyle{IEEEtran}
\bibliography{BibAbrv/IEEEabrv, BibAbrv/OSDAbrv, BibAbrv/LPAbrv, BibAbrv/SurveyAbrv, BibAbrv/ClassicAbrv, BibAbrv/MLAbrv, BibAbrv/MathAbrv,BibAbrv/NOMAAbrv}
	\vspace{-0.3em}

\end{document}